\documentclass{llncs}
\usepackage{llncsdoc}

\usepackage{graphicx}
\usepackage{times}
\usepackage{epsfig}
\usepackage[cmex10]{amsmath}
\usepackage{url}
\usepackage{amssymb}

\def\d3k{{\displaystyle {\rm d}{\bf k} \over \displaystyle (2\pi)^3}}
\def\dx{{\displaystyle {\rm d}{\bf x}}}
\def\Mpch{~h^{-1} {\rm Mpc}}

\newcommand{\cgal}{\texttt{CGAL}\ }
\newcommand{\mm}[1]      {\ifmmode{#1}\else{\mbox{\(#1\)}}\fi}
\newcommand{\Manifold}   {\mm{{\mathbb M}}}
\newcommand{\Rspace}     {\mm{{\mathbb R}}}
\newcommand{\Cspace}     {\mm{{\mathbb C}}}
\newcommand{\Betti}      {\mm{{\beta}}}
\newcommand{\Euler}      {\mm{\chi}}
\newcommand{\genus}      {\mm{G}}
\newcommand{\genusalt}   {\mm{g}}
\newcommand{\diff}       {\mm{\rm \,d}}
\newcommand{\Prob}       {\mm{\cal P}}
\newcommand{\Probability}{\mm{\rm Prob}}
\newcommand{\ProbDensity}{\mm{\cal P}}
\newcommand{\DMeasure}   {\mm{\cal D}}
\newcommand{\Matrix}     {\mm{\sf M}}
\newcommand{\ff}         {\mm{\bf f}}
\newcommand{\kk}         {\mm{\bf k}}

\newcommand{\xx}         {\mm{\bf x}}
\newcommand{\yy}         {\mm{\bf y}}
\newcommand{\XX}         {\mm{\bf X}}
\newcommand{\hatf}       {\mm{\hat f}}
\newcommand{\real}       {\mm{\rm re}}
\newcommand{\imaginary}  {\mm{\rm im}}
\newcommand{\imunit}     {\mm{\rm i}}

\begin{document}
\title{Alpha, Betti and the Megaparsec Universe:}
\subtitle{on the Topology of the Cosmic Web}
% Use 
\titlerunning{Alpha, Betti and the Megaparsec Universe} 
%for an abbreviated version of
% your contribution title if the original one is too long
\author{Rien van de Weygaert\inst{1} 
\and Gert Vegter\inst{2}
\and Herbert Edelsbrunner\inst{3}
\and Bernard J.T. Jones\inst{1}
\and Pratyush Pranav\inst{1}
\and Changbom Park\inst{4}
\and Wojciech A. Hellwing\inst{5}
\and\\ Bob Eldering\inst{2}
\and Nico Kruithof\inst{2}
\and E.G.P. (Patrick) Bos\inst{1}
\and Johan Hidding\inst{1}
\and \\ Job Feldbrugge\inst{1}
\and Eline ten Have\inst{6}
\and Matti van Engelen\inst{2}
\and \\ Manuel Caroli\inst{7}
\and Monique Teillaud\inst{7}}
% Use \authorrunning{Short Title} for an abbreviated version of
% your contribution title if the original one is too long
\institute{Kapteyn Astronomical Institute, University of Groningen, P.O. Box 800, 
  9700 AV Groningen, the Netherlands
\and 
Johann Bernoulli Institute for Mathematics and Computer Science, 
University of Groningen, P.O. Box 407, 9700 AK Groningen, the Netherlands
\and
IST Austria, Am Campus 1, 3400 Klosterneuburg, Austria
\and
School of Physics, Korea Institute for Advanced Study, Seoul 130-722, Korea
\and
Interdisciplinary Centre for Mathematical and Computational Modeling, 
University of Warsaw, ul. Pawinskiego 5a, 02-106 Warsaw, Poland
\and
Stratingh Institute for Chemistry, University of Groningen, 
Nijenborgh 4, 9747 AG Groningen, the Netherlands
\and
G\'eom\'etrica, INRIA Sophia Antipolis-M\'editerran\'ee, route des Lucioles, BP 93, 06902 Sophia Antipolis Cedex, France}

\maketitle

\begin{abstract}
We study the topology of the Megaparsec Cosmic Web in terms of the scale-dependent Betti numbers,
which formalize the topological information content of the cosmic mass distribution.
While the Betti numbers do not fully quantify topology, they extend the information 
beyond conventional cosmological studies of topology in terms of genus and Euler characteristic.
The richer information content of Betti numbers goes along the availability
of fast algorithms to compute them. 

For continuous density fields, we determine the scale-de\-pen\-dence of Betti numbers by 
invoking the cosmologically familiar filtration of sublevel or superlevel sets 
defined by density thresholds.
For the discrete galaxy distribution, however, the analysis is based on
the alpha shapes of the particles.
These simplicial complexes constitute an ordered sequence of nested subsets of
the Delaunay tessellation, a filtration defined by the scale parameter, $\alpha$.
As they are homotopy equivalent to the sublevel sets of the distance field, 
they are an excellent tool for assessing the topological structure of a discrete point distribution. 
In order to develop an intuitive understanding for the behavior of Betti numbers as a 
function of $\alpha$, and their relation to the morphological patterns in the Cosmic Web,
we first study them within the context of simple heuristic Voronoi clustering models.
These can be tuned to consist of specific 
morphological elements of the Cosmic Web, i.e.\ clusters, filaments, or sheets.
To elucidate the relative prominence of the various Betti numbers in different stages
of morphological evolution, we introduce the concept of alpha tracks. 

Subsequently, we address the topology of structures emerging in the 
standard LCDM scenario and in cosmological scenarios with alternative dark energy content.
The evolution of the Betti numbers is shown to reflect the hierarchical 
evolution of the Cosmic Web.
We also demonstrate that the scale-dependence of the Betti numbers yields  
a promising measure of cosmological parameters, with a potential
to help in determining the nature of dark energy and to probe primordial non-Gaussianities. 
We also discuss the expected Betti numbers as a function of 
the density threshold for superlevel sets of a Gaussian random field. 

Finally, we introduce the concept of persistent homology.
It measures scale levels of the mass distribution and allows us to separate small
from large scale features.
Within the context of the hierarchical cosmic structure formation,
persistence provides a natural formalism for a multiscale topology study of the Cosmic Web. 
\end{abstract}

%%%%%%%%%%%%%%%%%%%%%%%%%%%%%%%%%%%%%%%%%%%%%%%%%%%%%%%%%%%%%%%%%%%%%%%%%%%%%%%%%%%%
%%%%%%%%%%%%%%%%%%%%%%%%%%%%%%%%%%%%%%%%%%%%%%%%%%%%%%%%%%%%%%%%%%%%%%%%%%%%%%%%%%%%
\medskip
\section{Introduction: the Cosmic Web}
%%%%%%%%%%%%%%%%%%%%%%%%%%%%%%%%%%%%%%%%%%%%%%%%%%%%%%%%%%%%%%%%%%%%%%%%%%%%%%%%%%%%
%%%%%%%%%%%%%%%%%%%%%%%%%%%%%%%%%%%%%%%%%%%%%%%%%%%%%%%%%%%%%%%%%%%%%%%%%%%%%%%%%%%%

\begin{figure*}
 \vskip -1.00truecm
 \begin{center}
  \includegraphics[bb=0 0 510 477,width=0.98\textwidth]{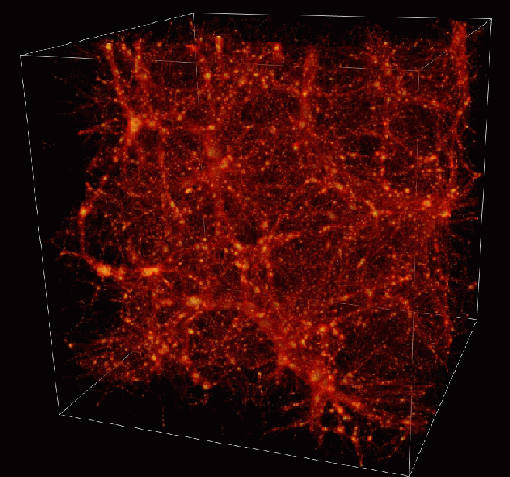} 
  \vskip -0.00truecm
  \caption{The Cosmic Web in an LCDM simulation.  Shown is the dark matter distribution
    at the current cosmological epoch in a $384^3$ particles $N$-body simulation
    of structure formation in a universe with a cosmological constant $\Lambda$, accounting for 
    an energy density $\Omega_\Lambda = 0.73$, 
    and cold dark matter, accounting for a total mass density of $\Omega_m = 0.23$.
    The box has a co-moving size of $80 \Mpch$ (i.e.\ a box co-expanding
    with the universe and having a current size of $ 80 \Mpch$). 
    Clearly visible is the intriguing network of filaments and high-density cluster nodes
    surrounding low-density voids.  The mass distribution has a distinct multiscale character, 
    marked by clumps over a wide range of mass scales, reflecting the hierarchical evolution of 
    the cosmic mass distribution. Figure courtesy of Bartosz Borucki.}
   \vskip -1.0truecm
  \label{fig:lcdmsimul}
 \end{center}
\end{figure*}

% no \IEEEPARstart
\noindent The large scale distribution of matter revealed by galaxy surveys features a complex 
network of interconnected filamentary galaxy associations. This network, which has become known as 
the {\it Cosmic Web} \cite{bondweb1996}, contains structures from a few megaparsecs\footnote{The main 
  measure of length in astronomy is the parsec.
  Technically, $1 \mbox{\rm pc}$ is the distance at which we would 
  see the distance Earth-Sun at an angle of 1 arcsec. It is equal to
  $3.262\,\,\mbox{\rm lightyears} = 3.086 \times 10^{13} \mbox{\rm km}$.
  Cosmological distances are substantially larger, so that a megaparsec,
  with $1 {\rm Mpc} = 10^6\,{\rm pc}$, is the common unit of distance.}
up to tens and even hundreds of megaparsecs of size. Galaxies and mass exist in a wispy web-like spatial arrangement 
consisting of dense compact clusters, elongated filaments, and sheet-like walls, amidst large near-empty 
voids, with similar patterns existing at earlier epochs, albeit over smaller scales;
see Figure~\ref{fig:lcdmsimul} for an illustration of a simulated cosmic mass distribution
in the standard LCDM cosmology\footnote{Currently, the LCDM cosmological scenario
  is the standard - or, ``concordance'' - cosmological scenario.
  It seems to be in agreement with a truly impressive amount of observational 
  evidence, although there are also some minor deficiencies. The ``L'' stands for $\Lambda$, the cosmological 
  constant that dominates the dynamics of our Universe, causes its expansion to accelerate and represents 
  in the order of about $73\%$ of its energy content. ``CDM'' indicates the Cold Dark Matter content 
  of the Universe. Invisible, it appears to form the major fraction of gravitating matter in the Universe, 
  representing some $23\%$ of the cosmic energy density as opposed to the mere $4.4\%$ that we find 
  in the normal baryonic matter we, and the planets and stars, consist of.} in a box of size $80 \Mpch$ 
\footnote{The Universe expands according to Hubble's law, stating that the recession velocity 
  of a galaxy is linearly proportional to its distance: $v = {\rm H} r$. The Hubble parameter, ${\rm H}$, quantifies 
  the expansion rate of the Universe, and is commonly expressed in units of km/s/Mpc. Its present value, the ``Hubble constant'', is 
  estimated to be ${\rm H}_0 \approx 71\,\,{\rm km/s/Mpc}$. Quite often, its value is expressed by a dimensionless 
  number, $h$, that specifies the Hubble parameter in units of $100\,\,{\rm km/s/Mpc}$.}.
The hierarchical nature of this mass distribution, marked by substructure over a wide range of scales and 
densities, has been clearly demonstrated \cite{weybond2008b}. Its appearance has been most dramatically 
illustrated by the recently produced maps of the nearby cosmos, the 2dFGRS, the SDSS and the 2MASS redshift 
surveys \cite{colless2003,tegmark2004,huchra2005} \footnote{Because of the expansion of the Universe, 
  any observed cosmic object will have its light shifted redward : its \emph{redshift}, $z$.
  According to Hubble's law, the redshift $z$ is directly proportional to the distance $r$ of the object,
  for $z \ll 1$: $cz={\rm H}r$, in which the constant ${\rm H}$ is the Hubble parameter. Because it is extremely 
  cumbersome to measure distances $r$ directly, cosmologists resort to the expansion of the Universe and 
  use $z$ as a distance measure. Because of the vast distances in the Universe, and the finite velocity of light, 
  the redshift $z$ of an object may also be seen as a measure of the time at which it emitted the observed radiation.}.

The vast Megaparsec Cosmic Web is one of the most striking examples of complex geometric patterns 
found in nature, and certainly the largest in terms of sheer size. Computer simulations suggest that 
the observed cellular patterns are a prominent and natural aspect of cosmic structure formation through 
gravitational instability \cite{peebles80}, the standard paradigm for the emergence of structure in 
our Universe \cite{weybond2008a,springmillen2005}. According to the {\it gravitational instability 
scenario}, cosmic structure grows from primordial density and velocity perturbations. These tiny 
primordial perturbations define a Gaussian density field and are fully characterized by the power 
spectrum; see Section~\ref{sec:gaussian}. 

%%%%%%%%%%%%%%%%%%%%%%%%%%%%%%%%%%%%%%%%%%%%%%%%%%%%%%%%%%%%%%%%%%%%%%%%%%%%%%%%%%%%%%
\subsection{Web Analysis}
%%%%%%%%%%%%%%%%%%%%%%%%%%%%%%%%%%%%%%%%%%%%%%%%%%%%%%%%%%%%%%%%%%%%%%%%%%%%%%%%%%%%%%

Over the past decades, we have seen many measures for characterizing different aspects
of the large scale cosmic structure: 
correlation functions (describing the $n$-point distribution),
Minkowski functionals and genus (characterizing the local and global curvature of isodensity surfaces),
multi-fractals (summarizing the statistical moments on various scales), and so on.  

Despite the multitude of descriptions, it has remained a major challenge to characterize the 
structure, geometry and topology of the Cosmic Web.
Many attempts to describe, let alone identify, the features and components of the Cosmic Web
have been of a rather heuristic nature.
The overwhelming complexity of both the individual structures as well as their connectivity,
the lack of structural symmetries, its intrinsic multiscale nature and the wide range of densities
in the cosmic matter distribution has prevented the use of simple and straightforward instruments. 

In the observational reality, galaxies are the main tracers of the Cosmic Web, and it is mainly through 
the measurement of the redshift distribution of galaxies that we have been able to map its structure.
Likewise, simulations of the evolving cosmic matter distribution are almost exclusively
based upon $N$-body particle computer 
calculations, involving a discrete representation of the features we seek to study. Both the galaxy distribution 
as well as the particles in an $N$-body simulation are examples of {\it spatial point processes} in that they 
are {\it discretely sampled} and have an {\it irregular spatial distribution}.

For furthering our understanding of the Cosmic Web, and to investigate its structure and dynamics, it is of prime importance 
to have access to a set of proper and objective analysis tools.
In this contribution, we address the topological and 
morphological analysis of the large scale galaxy distribution.
In particular, we introduce a new measure that is particularly 
suited to differentiating web-like structures: the homology of the distribution as measured by Betti numbers
and their persistence.

%%%%%%%%%%%%%%%%%%%%%%%%%%%%%%%%%%%%%%%%%%%%%%%%%%%%%%%%%%%%%%%%%%%%%%%%%%%%%%%%%%%%%%
\subsection{Genus and Minkowski Functionals}
%%%%%%%%%%%%%%%%%%%%%%%%%%%%%%%%%%%%%%%%%%%%%%%%%%%%%%%%%%%%%%%%%%%%%%%%%%%%%%%%%%%%%%

Gott and collaborators were the first to propose the use of the topology of the megaparsec structure 
of the Universe \cite{gott1986,hamilton1986},
to examine whether or not the initial fluctuations had been Gaussian.
They introduced the genus of isodensity surfaces of the cosmic mass distribution
as a quantitative characterization of cosmic topology in terms of the connectivity of these surfaces.
Being an intrinsic topology measure, the genus is relatively insensitive to systematic effects
such as non-linear gravitational evolution, galaxy biasing, and redshift-space distortions 
\cite{park2010}. This, in fact, is one of the most important advantages of using the genus in studying 
the patterns in the Megaparsec Universe. This allowed \cite{gott1986}, along with a series of subsequent 
studies, to conclude that the measured genus of the observed large scale structure is consistent with the 
predictions of cosmological scenarios whose initially Gaussian density and velocity fluctuations 
correspond to that of the observed power spectrum \cite{gott1989,park1992,moore1992,vogeley1994,park2005,gott2009,zhang2010}. 
Nonetheless, recent work has shown that on small scales, the observed genus is inconsistent with 
theoretical models \cite{choi2010}. On these scales the - not fully known - details of the galaxy formation process 
become important. 

Additional quantitative information on the morphology of the large scale distribution of galaxies
are the Minkowski functionals. Mecke et al. \cite{mecke1994} introduced them within a 
cosmological context as a more extensive characterization of the morphology and geometry of spatial patterns, 
after which Schmalzing et al. \cite{schmalzing1997,schmalzing1999} and others developed them
into a useful tool for studying cosmological datasets.
The Minkowski functionals measure the geometry and shape of manifolds,
providing a quantitative characterization of the level sets (the isodensity surfaces)
of the matter distribution at a given level of smoothing of the original data.
In a $d$-dimensional space, there are $d+1$ Minkowski functionals, and the Euler characteristic of isodensity 
surfaces is one of them.
In 3 dimensions, there are four Minkowski functionals: the volume enclosed by an isodensity surface,
the surface area, the integrated mean curvature, and the Euler characteristic (as in \eqref{eq:euler}).
From these, it is possible to generate nonparameteric shape descriptors such as 
the ``typical'' thickness, breadth and length of the structures involved at that level of smoothing \cite{sahni1998}. 
There are also algorithms for computing these quantities from discrete point sets \cite{aragon10},
and the technique has found wide applications outside of astronomy and physics.

What the Minkowski functionals do not tell about is the topology of the structure as described by the number of 
blobs, tunnels and voids that are present with changing scale. In that sense, the Betti numbers and their persistence
will provide a substantial extension of available topological information. Also, the related 
use of alpha shapes will provide effective methods of evaluating the Minkowski functionals.

%%%%%%%%%%%%%%%%%%%%%%%%%%%%%%%%%%%%%%%%%%%%%%%%%%%%%%%%%%%%%%%%%%%%%%%%%%%%%%%%%%%%%%
\subsection{Homology Groups}
\label{sec:introbetti}
%%%%%%%%%%%%%%%%%%%%%%%%%%%%%%%%%%%%%%%%%%%%%%%%%%%%%%%%%%%%%%%%%%%%%%%%%%%%%%%%%%%%%%

Here we advance the topological characterization of the Cosmic Web to a more complete description,
the homology of the distribution as measured by the scale-dependent Betti numbers of the sample.
While a full quantitative characterization 
of the topology of the cosmic mass distribution is not feasible, its homology
is an attractive compromise between detail and speed, providing a useful summary measurement of topology.
The ranks of the homology groups, known as the \emph{Betti numbers},
completely characterize the topology of orientable $2$-manifolds,
such as the isodensity surfaces used in earlier topological analyzes of the Cosmic Web.
For a $d$-dimensional manifold, we have $d+1$ homology groups, $H_p$, and correspondingly
$d+1$ Betti numbers, $\Betti_p$, for $0 \leq p \leq d$.
The Betti numbers may be considered as the number of $p$-dimensional holes.
In fact, homology groups can be seen as a useful definition of holes in the cosmic mass distribution.
They are fundamental quantities from which the Euler characteristic and genus are derived,
including those of the isodensity $2$-manifolds.
In that sense, they are a powerful generalization of genus analysis.

While the Euler characteristic and the Betti numbers give information about the connectivity of a manifold, 
two of the other three Minkowski functionals are sensitive to local manifold deformations,
and their topological information is limited.
The Betti numbers represent a more succinct as well as more informative characterization of the topology. 

Following this observation, we will argue and demonstrate that the scale-dependence of Betti numbers makes them 
particularly suited to analyze and differentiate web-like structures.
The Betti numbers measured as a function of scale are a 
sensitive discriminator of cosmic structure and can effectively reveal differences arising in alternative
cosmological models. Potentially, they may be exploited to infer information on important issues such as the nature of 
dark energy or even possible non-Gaussianities in the primordial density field, as will be corroborated 
by \cite{weygaert11}; see Section~\ref{sec:debetti}.

%%%%%%%%%%%%%%%%%%%%%%%%%%%%%%%%%%%%%%%%%%%%%%%%%%%%%%%%%%%%%%%%%%%%%%%%%%%%%%%%%%%%%%
\subsection{Alpha Shapes}
%%%%%%%%%%%%%%%%%%%%%%%%%%%%%%%%%%%%%%%%%%%%%%%%%%%%%%%%%%%%%%%%%%%%%%%%%%%%%%%%%%%%%%

Most of the topological studies in cosmology depend on some sort of user-controlled smoothing and related threshold 
to specify surfaces of which the topology may be determined. In cosmological studies, this usually concerns 
isodensity surfaces and their superlevel and sublevel sets defined on a Gaussian filter scale. 

A preferred alternative would be avoid filters and
to invoke a density field reconstruction that adapts itself to the galaxy distribution. For our purpose, 
the optimal technique would be that of the Delaunay Tessellation Field Estimator (DTFE) formalism 
\cite{schaapwey2000,weyschaap2009,cautun11}, which 
uses the local density and shape sensitivity of the Delaunay tessellation to recover a density field 
that retains the multiscale nature as well as the anisotropic nature of the sampled particle or galaxy 
distribution. This results in an adaptive and highly flexible representation of the underlying density 
field, which may then be used to assess its topological and singularity structure;
see e.g. \cite{zhang2010,aragon10,sousbie11a}. 

A closely related philosophy is to focus exclusively on the shape of the particle or galaxy 
distributions itself, and seek to analyze its topology without resorting to the isodensity surfaces and 
level sets of the corresponding density field.
This is precisely where \emph{alpha shapes} enter the stage. 
They are subsets of a Delaunay triangulation that describe the intuitive notion of the shape of 
a discrete point set. Introduced by Edelsbrunner and collaborators\cite{edelsbrunner1983}, they are one 
of the principal concepts from the field of Computational Topology \cite{dey1998,vegter2004,zomorodian2005}.

%%%%%%%%%%%%%%%%%%%%%%%%%%%%%%%%%%%%%%%%%%%%%%%%%%%%%%%%%%%%%%%%%%%%%%%%%%%%%%%%%%%%%%
\subsection{Outline}
%%%%%%%%%%%%%%%%%%%%%%%%%%%%%%%%%%%%%%%%%%%%%%%%%%%%%%%%%%%%%%%%%%%%%%%%%%%%%%%%%%%%%%

This review contains a report on our project to study the homology and topology of the Cosmic Web.
It is a major upgrade of an earlier report \cite{weygaert10}. 

Sections~\ref{sec:homology} and \ref{sec:alphashape} provide the mathematical background
for this review, consisting of a short presentation of homology and Betti numbers,
followed by a discussion of alpha shapes and the formalism to infer Betti numbers.

One of the intentions of our study is to understand the information content of
scale-dependent Betti numbers with respect to the corresponding geometric patterns observed
in the Cosmic Web, i.e.\ the relative prominence of clusters, filaments and walls.
To this end, we first investigate a number of Voronoi clustering models.
These models use the geometric structure of Voronoi tessellations as a skeleton of the 
cosmic mass distribution. In Section~\ref{sec:voronoimodel}, we describe these models,
and specify the class of Voronoi Element Models and Voronoi Kinematic Models.
Section~\ref{sec:voronoibetti} contains an extensive description of the analysis
of a set of Voronoi Element and Voronoi Kinematic Models.

Following the analysis of Voronoi clustering models, we turn to the homology analysis
in a few cosmological situations of current interest.
Section~\ref{sec:lcdmbetti} discusses the scale-dependent Betti numbers inferred
for the current standard theory of cosmic structure formation, LCDM. The results are obtained 
from a computer simulation of structure formation in this cosmological scenario.
The discussion in  Section~\ref{sec:lcdmbetti} connects the resulting homological
characteristics to the hierarchically evolving mass distribution.
In an attempt to investigate and exploit the sensitivity of Betti numbers to key
cosmological parameters, we apply our analysis to a set of cosmological scenarios 
with a different content of dark energy.
Section~\ref{sec:debetti} demonstrates that indeed homology may be a promising tool towards
inferring information on the dark energy content that dominates the dynamics of our Universe.
Equally compelling is the issue of the Betti numbers of superlevel or sublevel 
sets of Gaussian random fields.
To high accuracy, the initial density field out of which all structure in 
our Universe arose has had a Gaussian character. As reference point for any further assessment of 
the homology of the evolving cosmic mass distribution, we therefore need to evaluate the homology of 
Gaussian random fields. This is the subject of Section~\ref{sec:gaussian}, which focusses
on the expected values of the Betti numbers.
In addition, it will be a starting point for any related study 
looking for possible primordial non-Gaussian deviations. 

Having shown the potential of homology studies for cosmological purposes,
we discuss the prospects of assessing the persistence of cosmological density fields in Section~\ref{sec:persist}.
Persistence measures scale levels of the mass distribution and allows us to systematically separate
small from large scale features in the mass distribution.
It provides us with a rich language to study intrinsically multiscale 
distributions as those resulting from the hierarchically evolving structure in the Universe. 
Finally, Section~\ref{sec:conclusions} addresses future prospects and relates this review to other work.

%%%%%%%%%%%%%%%%%%%%%%%%%%%%%%%%%%%%%%%%%%%%%%%%%%%%%%%%%%%%%%%%%%%%%%%%%%%%%%%%%%%%%%
%%%%%%%%%%%%%%%%%%%%%%%%%%%%%%%%%%%%%%%%%%%%%%%%%%%%%%%%%%%%%%%%%%%%%%%%%%%%%%%%%%%%%%
\medskip
\section{Homology and Betti numbers}
\label{sec:homology}
%%%%%%%%%%%%%%%%%%%%%%%%%%%%%%%%%%%%%%%%%%%%%%%%%%%%%%%%%%%%%%%%%%%%%%%%%%%%%%%%%%%%%%
%%%%%%%%%%%%%%%%%%%%%%%%%%%%%%%%%%%%%%%%%%%%%%%%%%%%%%%%%%%%%%%%%%%%%%%%%%%%%%%%%%%%%%

Homology groups and Betti numbers are concepts from algebraic topology,
designed to quantify and compare topological spaces.  
They characterize the topology of a space in terms of the relationship
between the cycles and boundaries we find in the space\footnote{Assuming
  the space is given as a simplicial complex,
  a \emph{$p$-cycle} is a $p$-chain with empty boundary,
  where a \emph{$p$-chain}, $\gamma$, is a sum of $p$-simplices.
  The standard notation is $\gamma = \sum a_i \sigma_i$, 
  where the $\sigma_i$ are the $p$-simplices and the $a_i$ are the {\it coefficients}.
  For example, a 1-cycle is a closed loop of edges, or a finite union of such loops,
  and a 2-cycle is a closed surface, or a finite union of such surfaces.
  Adding two $p$-cycles, we get another $p$-cycle, and similar for the $p$-boundaries.
  Hence, we have a group of $p$-cycles and a group of $p$-boundaries.}.
For example, if the space is a $d$-dimensional manifold, $\Manifold$,
we have cycles and boundaries of dimension $p$ from $0$ to $d$.
Correspondingly, $\Manifold$ has one homology group $H_p (\Manifold)$
for each of $d+1$ dimensions, $0 \leq p \leq d$.
By taking into account that two cycles should be considered identical if
they differ by a boundary, one ends up with a group $H_p (\Manifold)$
whose elements are the equivalence classes of $p$-cycles\footnote{Correctly defined,
  the \emph{$p$-th homology group} is the $p$-th cycle group modulo the $p$-th boundary group.
  In algebraic terms, this is a quotient group.
  In intuitive terms, this says that two $p$-cycles are considered the same,
  or \emph{homologous}, if together they bound a $(p+1)$-chain or, equivalently,
  if they differ by a $p$-boundary.
  Indeed, we do not want to distinguish between two $1$-cycles of, say the torus,
  if they both go around the hole, differing only in the geometric paths
  they take to do so.}.
The \emph{rank} of the homology group $H_p (\Manifold)$ is the $p$-th \emph{Betti number},
$\Betti_p = \Betti_p (\Manifold)$.

In heuristic - and practical - terms, the Betti numbers count topological features
and can be considered as the number of $p$-dimensional holes.
When talking about a surface in $3$-dimensional space,
the zeroth Betti number, $\Betti_0$, counts the components,
the first Betti number, $\Betti_1$, counts the tunnels,
and the second Betti number, $\Betti_2$, counts the enclosed voids.
All other Betti numbers are zero.
Examples of spaces with Betti numbers which are of interest to us
are the sublevel and superlevel density set of the cosmic mass distribution,
i.e.\ the regions whose density is smaller or greater than a specified threshold level.

%%%%%%%%%%%%%%%%%%%%%%%%%%%%%%%%%%%%%%%%%%%%%%%%%%%%%%%%%%%%%%%%%%%%%%%%%%%%%%%%%%%%%%%%%
\subsection{Genus and Euler Characteristic} 
\label{sec:genus}
%%%%%%%%%%%%%%%%%%%%%%%%%%%%%%%%%%%%%%%%%%%%%%%%%%%%%%%%%%%%%%%%%%%%%%%%%%%%%%%%%%%%%%%%%

Numerous cosmological studies have considered the \emph{genus} of the isodensity surfaces
defined by the megaparsec galaxy distribution \cite{gott1986,hamilton1986,hoyle2002},
which specifies the number of handles defining the surface.
More formally, it is the maximum number of disjoint closed curves such that cutting
along these curves does not increase the number of components.
The genus has a simple relation to the Euler characteristic, $\Euler$, of the isodensity surface.
Consider a $3$-manifold subset $\Manifold$ of the Universe and its boundary,
$\partial \Manifold$, which is a $2$-manifold.
With $\partial \Manifold$ consisting of $c = \Betti_0 (\partial \Manifold)$ components,
the Gauss-Bonnet Theorem states that the genus of the surface is given by
\begin{eqnarray}
  \genus  &=&  c - \frac{1}{2} \Euler (\partial \Manifold) ,
\label{eq:genus}
\end{eqnarray}
where the Euler characteristic $\Euler (\partial \Manifold)$ is
the integrated Gaussian curvature of the surface 
\begin{eqnarray}
  \Euler (\partial \Manifold)  &=&  \frac{1}{2 \pi} \oint_x \frac{\diff x}{R_1 (x) R_2 (x) } . 
  \label{eq:euler}
\end{eqnarray}
Here $R_1 (x)$ and $R_2 (x)$ are the principal radii of curvature at the point $x$ of the surface.
The integral of the Gaussian curvature is invariant under continuous deformation of the surface:
perhaps one of the most surprising results in differential geometry.

The Euler characteristic of the surface can also be expressed in terms of
its Betti numbers, namely, $\chi (\partial \Manifold)$ is equal to
$\Betti_0 (\partial \Manifold)$ minus $\Betti_1 (\partial \Manifold)$
plus $\Betti_2 (\partial \Manifold)$.
This is implied by the Euler-Poincar\'{e} Formula, which we will discuss shortly,
after introducing triangulation of spaces.
Combining these two equations for the Euler characteristic,
we get a fundamental relationship between differential geometry and algebraic topology.
Returning to the $3$-dimensional subset, $\Manifold$, of the Universe,
its Euler characteristic is
\begin{eqnarray}
  \Euler (\Manifold)  &=&  \Betti_0 (\Manifold) - \Betti_1 (\Manifold)
                         + \Betti_2 (\Manifold) - \Betti_3 (\Manifold) .
\end{eqnarray}
Whenever $\Manifold$ is a non-exhaustive subset of the connected Universe,
its third Betti number vanishes, $\Betti_3 (\Manifold) = 0$,
and its boundary is necessarily non-empty, namely a $2$-manifold without boundary.
As mentioned above, the Euler characteristic of $\partial \Manifold$ is the
alternating sum of Betti numbers,
where $\Betti_0 (\partial \Manifold)$ is the number of surface components,
$\Betti_1 (\partial \Manifold)$ is twice the genus, and
$\Betti_2 (\partial \Manifold) = \Betti_0 (\partial \Manifold)$.
Assuming the Universe is connected like the $3$-sphere, we can use Alexander duality
and the Mayer-Vietoris sequence to establish a direct relation between the Betti numbers of the 
$3$-manifold with boundary, $\Manifold$, and those of the $2$-manifold without boundary,
$\partial \Manifold$; see e.g.\ \cite{edelsbrunner2009}: 
\begin{eqnarray}
  \Betti_0 (\partial \Manifold)  &=&  \Betti_2 (\partial \Manifold) 
                                ~~=~~ \Betti_0 (\Manifold) + \Betti_2 (\Manifold) , \\
  \Betti_1 (\partial \Manifold)  &=&  2 \Betti_1 (\Manifold) .
\end{eqnarray}
From this, we infer that the Euler characteristic of the boundary
is directly proportional to the Euler characteristic of the $3$-manifold:
\begin{eqnarray}
  \Euler (\partial \Manifold)  &=&  \Betti_0 (\partial \Manifold) - \Betti_1 (\partial \Manifold)
                                  + \Betti_2 (\partial \Manifold)                      \\
                               &=&  2 \Euler (\Manifold) .
\end{eqnarray}
In the cosmologically interesting situation in which $\partial \Manifold$
is the isodensity surface of either density superlevel or sublevel sets,
we find a relation between the genus\footnote{For consistency,
  it is important to note that the definition in previous 
  topology studies in cosmology \cite{gott1986,hamilton1986} slightly differs from this.
  The genus, $\genusalt$, in these studies has been defined as the number of holes minus the 
  number of connected regions: $\genusalt = \genus - c$.
  Here, we will refer to $g$ as the \emph{reduced genus}.}
of the surface and the Betti numbers of the enclosed manifold:
\begin{eqnarray}
  \genus  &=&  c - \frac{1}{2} \Euler (\partial \Manifold)                                             \\
          &=&  c - \Euler (\Manifold)                                                                  \\
          &=&  c - \left( \Betti_0 (\Manifold) - \Betti_1 (\Manifold) + \Betti_2 (\Manifold) \right) .
               \label{eqn:genus-Betti}
\end{eqnarray}
In the analysis described in this paper, we will restrict ourselves to the 
three Betti numbers, $\Betti_0$, $\Betti_1$, and $\Betti_2$,
of the $3$-manifolds with boundary defined by the cosmic mass distribution.

%%%%%%%%%%%%%%%%%%%%%%%%%%%%%%%%%%%%%%%%%%%%%%%%%%%%%%%%%%%%%%%%%%%%%%%%%%%%%%%%%%%%%%
\subsection{Triangulated Spaces} 
%%%%%%%%%%%%%%%%%%%%%%%%%%%%%%%%%%%%%%%%%%%%%%%%%%%%%%%%%%%%%%%%%%%%%%%%%%%%%%%%%%%%%%

A practical simplification occurs when we represent a space by a \emph{triangulation},
which is a simplicial complex that retains the topological properties of the space.
These are topological spaces assembled from vertices, edges, triangles, tetrahedra,
and possibly higher-dimensional simplices\footnote{Technically,
  we require that the union of simplices in the simplicial complex
  is \emph{homeomorphic} to the space it represents,
  which means that there is a bijective map between the two sets that
  is continuous and whose inverse is continuous.
  As an example, consider the boundary of the octahedron,
  consisting of $6$ vertices, $12$ edges, and $8$ triangles, which
  forms a triangulation of the $2$-dimensional sphere.}.
In this situation, homology is defined as described earlier, by comparing
chains that are cycles with chains that are boundaries.
The availability of simplices has a number of advantages,
including the existence of fast algorithms to compute homology.
Here, we focus on the connection between the number of simplices
in the simplicial complex and the Betti numbers of the space.

Suppose $\Manifold$ is a $d$-dimensional manifold,
and $K$ is a triangulation of $\Manifold$.
Write $n_p$ for the number of $p$-dimensional simplices in $K$.
For example, $n_0$ is the number of vertices, $n_1$ is the number of edges, and so on.
The \emph{Euler characteristic} of $K$ is defined as the alternating sum of these numbers:
\begin{eqnarray}
  \Euler (K)  &=&  \sum_{p=0}^d (-1)^p n_p .
\end{eqnarray}
This is the $d$-dimensional generalization of the classical Euler characteristic of a polytope:
\begin{eqnarray}
  \Euler  &=&  \mbox{\rm \#vertices} - \mbox{\rm \#edges} + \mbox{\rm \#faces} .
\end{eqnarray}
For a convex polytope, Euler's Formula states that $\Euler = 2$. 
The \emph{Euler-Poincar\'{e} Formula} is a far-reaching generalization of this relation.
To state this generalization, we first note that homology is independent
of the choice of triangulation, and so are the Betti numbers and the Euler characteristic.
The Euler-Poincar\'{e} Formula says that the Euler characteristic of a
triangulation, which is the alternating sum of simplex numbers,
is equal to the alternating sum of Betti numbers of the triangulated space:
\begin{eqnarray}
  \Euler (K)  &=&  \sum_{p=0}^d (-1)^p \Betti_p (\Manifold) .
\end{eqnarray}
Coming back to the case of a convex polytope in $3$-dimensional space,
its faces decompose the boundary, which is homeomorphic to the $2$-dimensional sphere.
We may further decompose the faces into triangles, if necessary, but this makes no difference here.
Since this is true for all convex polytopes, their faces are but different
triangulations of this same sphere, so the alternating sums must be the same.
They are all equal to $2$ because this is the Euler characteristic of the
$2$-dimensional sphere.

\begin{figure*}[t]
 \begin{center}
  \vskip -0.0truecm
  \includegraphics[bb=0 0 1277 786,width=0.98\textwidth]{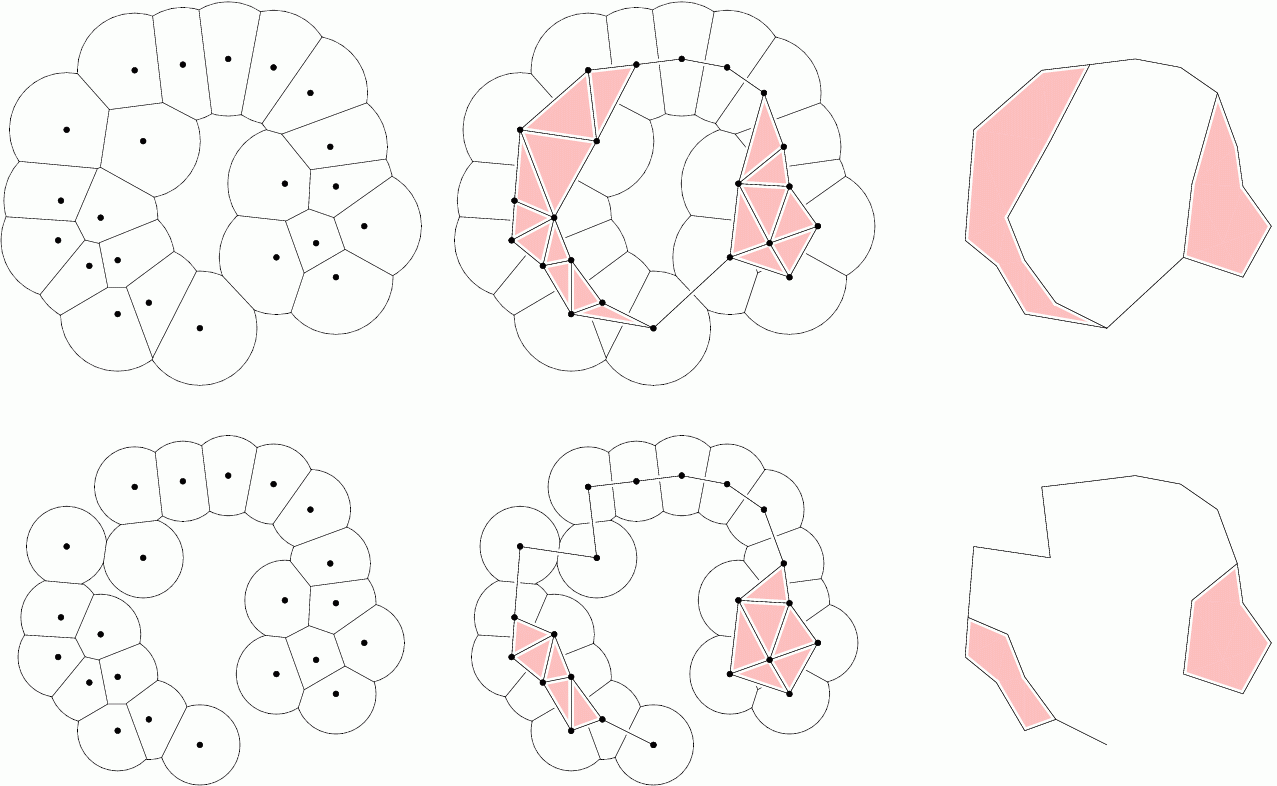} 
%  \centerline{\epsfig{figure=Rien_Weygaert_cosmoalphabetti_LNCS_Fig2.eps,width=0.98\textwidth}}
  \vskip 0.25truecm
  \caption{Illustration of alpha shapes.  For two different values of $\alpha$,
    we show the relation between the $2$-dimensional point distribution, the value of $\alpha$,
    and the resulting alpha shape.  Around each point in the sample, we draw a circle of radius $\alpha$.
    The outline of the corresponding Voronoi tessellation is indicated by the edges (left).
    All Delaunay simplices \emph{dual} to the decomposition of the union of disks by the Voronoi
    polygons are shown in black (center).  The final resulting alpha shape is shown on the right.
    Top: large $\alpha$ value. Bottom: small $\alpha$ value.}
  \label{fig:alphashape}
 \end{center}
\end{figure*}

%%%%%%%%%%%%%%%%%%%%%%%%%%%%%%%%%%%%%%%%%%%%%%%%%%%%%%%%%%%%%%%%%%%%%%%%%%%%%%%%%%%%%%
\subsection{Homology of a Filtration}
%%%%%%%%%%%%%%%%%%%%%%%%%%%%%%%%%%%%%%%%%%%%%%%%%%%%%%%%%%%%%%%%%%%%%%%%%%%%%%%%%%%%%%

For the assessment of the topology of a mass or point distribution,
a rich source of information is the topological structure of a filtration.
Given a space $\Manifold$, a \emph{filtration} is a nested sequence of subspaces:
\begin{equation}
  \emptyset  = \Manifold_0 \subseteq \Manifold_1 \subseteq \ldots \subseteq \Manifold_m = \Manifold .
\end{equation}
The nature of the filtrations depends, amongst others, on the representation of the mass distribution. 
When representing the mass distribution by a continuous density field, $f({\xx})$,
a common practice is to study the sublevel or superlevel sets of the field smoothed on a scale $R_s$:
\begin{eqnarray}
  f_s (\xx)  &=&  \int f(\yy) W_s(\yy-\xx) \diff \yy ,
\end{eqnarray}
where $W_s({\bf x}-{\bf y})$ is the smoothing kernel.
The \emph{sublevel sets} of this field are defined as the regions
\begin{eqnarray}
  \Manifold_\nu  &=&  \left\{ \xx \in \Manifold \mid f_s (\xx) \in (-\infty, f_{\nu}] \right\} \\
                 &=&  f_s^{-1} (-\infty, f_{\nu}] . 
\end{eqnarray}
In other words, they are the regions where the smoothed density
is less than or equal to the threshold value
$f_{\nu} = \nu \sigma_0$, with $\sigma_0$ the dispersion of the density field.
When addressing the topology of the primordial 
Gaussian density field, as in Section~\ref{sec:gaussian},
the analysis will be based on the filtration consisting of superlevel sets.

The representation of the mass by a discrete particle or galaxy distribution leads 
to an alternative strategy of filtration, this time in terms of simplicial complexes
generated by the particle distribution.
In the case of simplicial complexes that are homotopy equivalent to the sublevel sets
of either the density or the distance function field,
the computation of the homological characteristics of the field is considerably facilitated.
Often it is also much easier to visualize the somewhat 
abstract notion of homology by means of these simplicial complexes.

In our study, we will concentrate on \emph{alpha shapes} of a point set, which are subsets of 
the corresponding Delaunay triangulation; see Section~\ref{sec:alphashape}.
The alpha shapes are homotopy equivalent to the sublevel sets of the distance field
defined by the point distribution, and they constitute a nested sequence of simplicial complexes
that forms a topologically useful filtration of the Delaunay triangulation. 
In the following sections, we will extensively discuss the use of alpha shapes to assess the 
homology of cosmological particle and galaxy distributions.

%%%%%%%%%%%%%%%%%%%%%%%%%%%%%%%%%%%%%%%%%%%%%%%%%%%%%%%%%%%%%%%%%%%%%%%%%%%%%%%%%%
%%%%%%%%%%%%%%%%%%%%%%%%%%%%%%%%%%%%%%%%%%%%%%%%%%%%%%%%%%%%%%%%%%%%%%%%%%%%%%%%%%
\medskip
\section{Alpha Shapes}
\label{sec:alphashape}
%%%%%%%%%%%%%%%%%%%%%%%%%%%%%%%%%%%%%%%%%%%%%%%%%%%%%%%%%%%%%%%%%%%%%%%%%%%%%%%%%%
%%%%%%%%%%%%%%%%%%%%%%%%%%%%%%%%%%%%%%%%%%%%%%%%%%%%%%%%%%%%%%%%%%%%%%%%%%%%%%%%%%

One of the key concepts in the field of Computational Topology are \emph{alpha shapes},
as introduced by Edelsbrunner and
collaborators \cite{edelsbrunner1983,mueckephd1993,edelsbrunner1994};
see \cite{edelsbrunner2009} for a recent review.
They generalize the convex hull of a point set and are concrete geometric objects
that are uniquely defined for a particular point set and a real value $\alpha$.
For their definition, we look at the union of balls of radius $\alpha$ centered on 
the points in the set, and its decomposition by the corresponding Voronoi tessellation;
see the left diagrams in Figure~\ref{fig:alphashape}.
The \emph{alpha complex} consists of all Delaunay simplices that record
the subsets of Voronoi cells that have a non-empty common intersection
within this union of balls; see the center diagrams of Figure~\ref{fig:alphashape}. 
The \emph{alpha shape} is the union of simplices in the alpha complex;
see the right diagrams of Figure~\ref{fig:alphashape}.

Alpha shapes reflect the topological structure of a point distribution on a scale 
parameterized by the real number $\alpha$.
The ordered set of alpha shapes constitute a filtration of the Delaunay tessellation.
The link between alpha shapes and the homology of a point distribution can be appreciated from 
the fact that tunnels will be formed when, at a certain value of $\alpha$,
an edge is added between two vertices that were already connected,
which increases the first Betti number.
When new triangles are added, the tunnel may be filled, which decreases the first Betti number.
More about this process of growing and shrinking Betti numbers in Section~\ref{sec:persist},
where we discuss the persistence of tunnels and other topological features.

\begin{figure*}[t]
 \begin{center}
  \vskip -0.0truecm
  \includegraphics[bb=0 0 1220 816,width=0.98\textwidth]{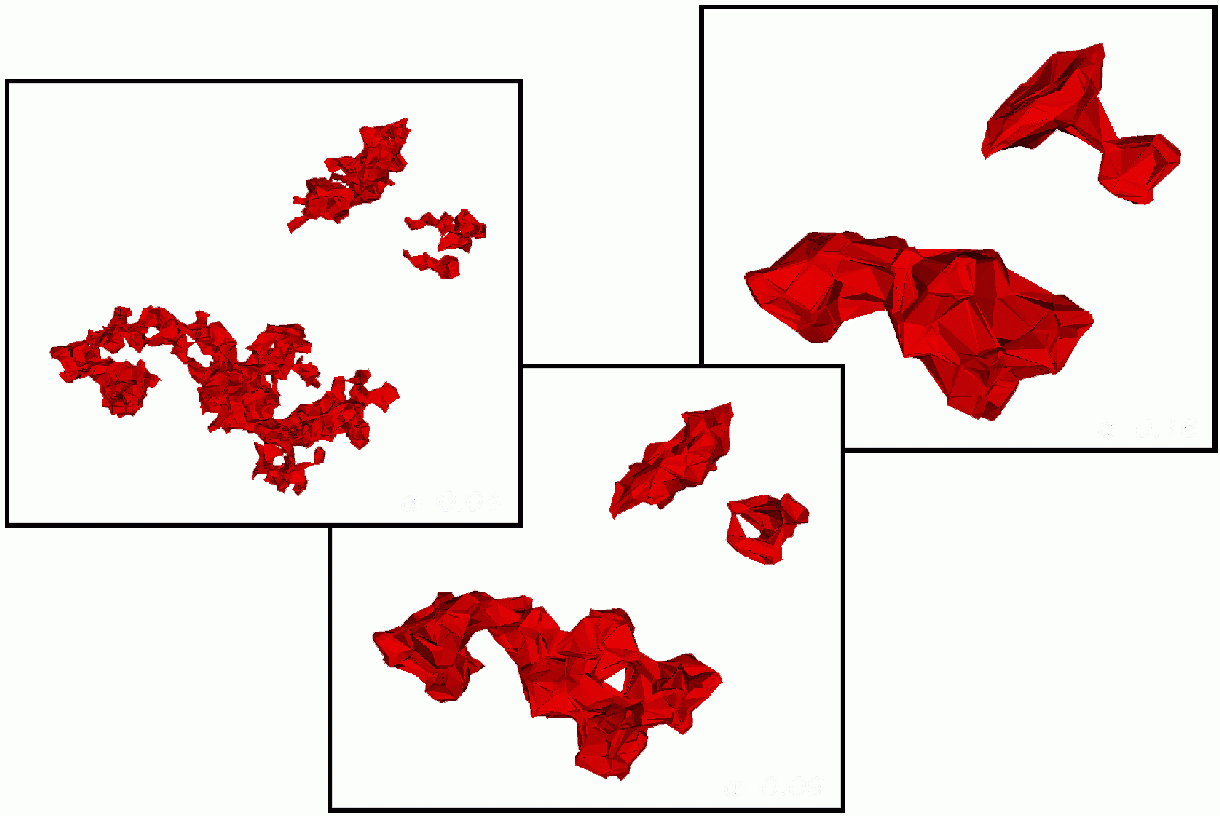} 
  \vskip 0.0truecm
  \caption{Illustration of $3$-dimensional alpha shapes for three different values
    of $\alpha = 0.001, 0.005, 0.01$.  Clearly visible are the differences in connectivity
    and topological structure of the different alpha shapes.}
  \label{fig:alpha3d}
 \end{center}
\end{figure*}
Connections of alpha shapes to diverse areas in the sciences and engineering have developed,
including to pattern recognition, digital shape sampling and processing,
and structural molecular biology \cite{edelsbrunner2009}. 
Applications of alpha shapes have as yet focussed on biological systems,
where they have been used in the characterization of the topology and structure of macromolecules.
The work by Liang and collaborators \cite{edelsbrunner1998,liang1998a,liang1998b,liang1998c}
uses alpha shapes and Betti numbers to assess the voids and pockets in an effort to classify
complex protein structures, a highly challenging task given the tens of thousands
of protein families involving thousands of different folds.
Given the interest in the topology of the cosmic
mass distribution \cite{gott1986,mecke1994,schmalzing1997,schmalzing1999},
it is evident that alpha shapes also provide a highly interesting tool for studying 
the topology of the galaxy distribution and the particles in $N$-body simulations
of cosmic structure formation.
Directly connected to the topology of the point distribution itself, 
it avoids the need of any user-defined filter kernels. 

%%%%%%%%%%%%%%%%%%%%%%%%%%%%%%%%%%%%%%%%%%%%%%%%%%%%%%%%%%%%%%%%%%%%%%%%%%%%%%%%%%%%%%
\subsection{Definition}
%%%%%%%%%%%%%%%%%%%%%%%%%%%%%%%%%%%%%%%%%%%%%%%%%%%%%%%%%%%%%%%%%%%%%%%%%%%%%%%%%%%%%%

Figure~\ref{fig:alphashape} provides an impression of the concept by
illustrating the process of defining the alpha shape, for two different values of $\alpha$.
If we have a finite point set, $S$, in $3$-dimensional space and its Delaunay triangulation,
we may identify all simplices -- vertices, edges, triangles, tetrahedra -- in the triangulation.
For a given non-negative value of $\alpha$, the \emph{alpha complex} consists of all simplices
in the Delaunay triangulation that have an empty circumsphere with radius less than or equal to $\alpha$.
Here ``empty'' means that the open ball bounded by the sphere does not include any points of $S$.
For an extreme value $\alpha = 0$, the alpha complex merely consists of the vertices of the point set.
There is also a smallest value, $\alpha_{\max}$, such that
for $\alpha \geq \alpha_{\max}$, the alpha complex is the Delaunay triangulation
and the alpha shape is the convex hull of the point set. 

As mentioned earlier, the alpha shape is the union of all simplices in the alpha complex.
It is a polytope in a fairly general sense: it can be concave and even disconnected.
Its components can be three-dimensional clumps of tetrahedra,
two-dimensional patches of triangles, one-dimensional strings of edges, and collections of isolated points,
as well as combinations of these four types.
The set of real numbers leads to a family of shapes capturing the intuitive notion
of the overall versus fine shape of a point set.
Starting from the convex hull gradually decreasing $\alpha$,
the shape of the point set gradually shrinks and starts to develop enclosed voids.
These voids may join to form tunnels and larger voids.
For negative $\alpha$, the alpha shape is empty.
An intuitive feel for the evolution of the topological structure
may be obtained from the three different alpha shapes of the same point set in
$3$-dimensional space shown in Figure~\ref{fig:alpha3d}. 

It is important to realize that alpha shapes are not triangulations of the
union of balls in the technical sense.
Instead, they are simplicial complexes that are homotopy equivalent to the
corresponding union of balls with radius $\alpha$.
While this is a nontrivial observation that follows from the Nerve Theorem
\cite{edelsbrunner2010}, it is weaker than being homeomorphic, which would
be necessary for being a triangulation.
Nevertheless, the homotopy equivalence implies  that the alpha shape and
the corresponding union of balls have the same Betti numbers.

\begin{figure*}
 \begin{center}
  \vskip -1.0truecm
  \mbox{\hskip -0.5truecm\includegraphics[bb=0 0 385 534,height=17.5truecm]{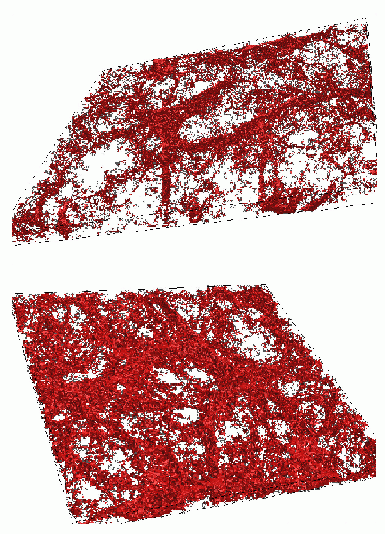}} 
  \vskip 0.15truecm
  \caption{Examples of alpha shapes of the LCDM GIF simulation.  Shown are central slices through 
    the complete alpha shape for two different values of $\alpha$, viewed from different angles.
    The sensitivity to the structure and topology of the matter distribution in the Cosmic Web
    is clearly visible when comparing the lower value of $\alpha$ in the top panel
    with the higher value of $\alpha$ in the bottom panel.}
  \label{fig:gifalphashape}
 \end{center}
\end{figure*}

Although the alpha shape is defined for all real numbers $\alpha$,
there are only a finite number of different alpha shapes for any finite point set.
In other words, the alpha shape process is never continuous: 
it proceeds discretely with increasing $\alpha$, marked by the addition of new Delaunay simplices
once $\alpha$ exceeds the corresponding threshold.

%%%%%%%%%%%%%%%%%%%%%%%%%%%%%%%%%%%%%%%%%%%%%%%%%%%%%%%%%%%%%%%%%%%%%%%%%%%%%%%%%%%%%%
\subsection{Computing Betti Numbers}
%%%%%%%%%%%%%%%%%%%%%%%%%%%%%%%%%%%%%%%%%%%%%%%%%%%%%%%%%%%%%%%%%%%%%%%%%%%%%%%%%%%%%%

Following the description above, one may find that alpha shapes are intimately related
to the topology of a point set. 
Indeed, they form a direct way of characterizing the topology of a point distribution.
The complete description of its homology in terms of Betti numbers may therefore
be inferred from the alpha shapes.

For simplicial complexes, like Delaunay tessellations and alpha complexes,
the Betti numbers can be defined on the basis of the $p$-simplices.
We illustrate this for a simplicial complex in $3$-dimensional space.
Cycling through all its simplices, we base the calculation on the
following straightforward considerations.
When a vertex is added to the alpha complex,
a new component is created and $\Betti_0$ is increased by $1$.
Similarly, if an edge is added, it connects two vertices, which either belong to the same or to
different components of the current complex.
In the former case, the edge creates a new tunnel, so $\Betti_1$ is increased by $1$.
In the latter case, two components get connected into one, so $\Betti_0$ is decreased by $1$.
If a triangle is added, it either completes a void or it closes a tunnel.
In the former case, $\Betti_2$ is increased by $1$,
and in the latter case, $\Betti_1$ is decreased by $1$.
Finally, when a tetrahedron is added, a void is filled, so $\Betti_2$ is lowered by $1$. 
Following this procedure, the algorithm has to include a technique for determining
whether a $p$-simplex belongs to a $p$-cycle.
For vertices and tetrahedra, this is rather trivial.
On the other hand, for edges and triangles, we use a somewhat more elaborate procedure,
involving the classical computer science concept of
a union-find data structure \cite{delfinado1993}.

Turning our attention to software that implements these algorithmic ideas,
we resort to the Computational Geometry Algorithms Library, \cgal\footnote{\cgal
  is a \texttt{C++} library of algorithms and data structures for computational geometry,
  see \url{www.cgal.org}.}.
Within this context, Caroli and Teillaud recently developed an efficient code
for the calculation of two-dimensional and three-dimensional alpha shapes in periodic spaces.
We use their software for the computation of the alpha shapes of our cosmological models.
In the first stage of our project, which concerns the analysis of Voronoi clustering models,
we computed the Betti numbers of alpha shapes with a code developed within our own project.
Later, for the analysis of the cosmological LCDM models, we were provided with
an optimized code written by Manuel Caroli.

%%%%%%%%%%%%%%%%%%%%%%%%%%%%%%%%%%%%%%%%%%%%%%%%%%%%%%%%%%%%%%%%%%%%%%%%%%%%%%%%%%%%%%
\subsection{Alpha Shapes of the Cosmic Web}
%%%%%%%%%%%%%%%%%%%%%%%%%%%%%%%%%%%%%%%%%%%%%%%%%%%%%%%%%%%%%%%%%%%%%%%%%%%%%%%%%%%%%%

In a recent study, Vegter et al.\ computed the alpha shapes for a set of GIF simulations 
of cosmic structure formation \cite{eldering2006}.
It concerns a $256^3$ particles GIF $N$-body simulation, encompassing
a LCDM ($\Omega_m = 0.3$, $\Omega_{\Lambda} = 0.7$, ${\rm H}_0 = 70\,\,{\rm km/s/Mpc}$)
density field within a (periodic) cubic box with length $141 \Mpch$  
and produced by means of an adaptive ${\rm P^3M}$ $N$-body code \cite{kauffmann1999}. 

Figure~\ref{fig:gifalphashape} illustrates the alpha shapes for two different values of $\alpha$, 
by showing sections through the GIF simulation.
The top panel uses a small value of $\alpha$, the bottom uses a high value.
The intricacy of the web-like patterns is very nicely followed.
The top configuration highlights the interior of filamentary and sheet-like features,
and reveals the interconnection between these structural elements. 
The bottom configuration covers an evidently larger volume, which it does by connecting 
finer features in the Cosmic Web. Noteworthy are the tenuous filamentary and 
planar extensions into the interior of the voids.  

These images testify of the potential power of alpha shapes in analyzing the
web-like cosmic matter distribution, in identifying its morphological elements,
their connections and, in particular, their hierarchical character. 
However, to understand and properly interpret the topological information contained in these 
images, we need first to assess their behavior in simpler yet similar circumstances.
To this end, we introduce a set of heuristic spatial matter distributions,
Voronoi clustering models.

%%%%%%%%%%%%%%%%%%%%%%%%%%%%%%%%%%%%%%%%%%%%%%%%%%%%%%%%%%%%%%%%%%%%%%%%%%%%%%%%%%
%%%%%%%%%%%%%%%%%%%%%%%%%%%%%%%%%%%%%%%%%%%%%%%%%%%%%%%%%%%%%%%%%%%%%%%%%%%%%%%%%%
\medskip
\section{Voronoi Clustering Models}
\label{sec:voronoimodel}
%%%%%%%%%%%%%%%%%%%%%%%%%%%%%%%%%%%%%%%%%%%%%%%%%%%%%%%%%%%%%%%%%%%%%%%%%%%%%%%%%%
%%%%%%%%%%%%%%%%%%%%%%%%%%%%%%%%%%%%%%%%%%%%%%%%%%%%%%%%%%%%%%%%%%%%%%%%%%%%%%%%%%

In this section, we introduce a class of heuristic models for cellular distributions
of matter using Voronoi tessellations as scaffolds,
know as \emph{Voronoi clustering models} \cite{weyicke1989,weygaert1991,weygaert2007}. 
They are well suited to model the large scale clustering of the morphological elements
of the Cosmic Web, defined by the stochastic yet non-Poissonian geometrical
distribution of the matter and the related galaxy population
forming \emph{walls}, \emph{filaments}, and \emph{clusters}.

The small-scale distribution of galaxies within the various components
of the cosmic skeleton involves the complicated details of highly nonlinear
interactions of the gravitating matter.
This aspect may be provided by elaborate physical models and/or $N$-body computer simulations,
but this would distract from the purpose of the model and destroy its conceptual simplicity.
In the Voronoi models, we complement a geometrically fixed Voronoi tessellation
defined by a small set of nuclei with a heuristic prescription for the location
of particles or model galaxies within the tessellation. 
We distinguish two complementary approaches: the \emph{Voronoi element}
and the \emph{Voronoi evolution models}.
Both are obtained by moving an initially random distribution of $N$ particles
toward the faces, lines, and nodes of the Voronoi tessellation.
The Voronoi element models do this by a heuristic and user-specified mixture of projections
onto the various geometric elements of the tessellation.
The Voronoi evolution models accomplish this via a gradual motion of the galaxies
from their initial, random locations towards the boundaries of the cells.

\begin{figure*}[t]
  \centering
     \vskip -0.5truecm
      \mbox{\hskip -0.5truecm\includegraphics[bb=0 0 485 163,width=0.98\textwidth]{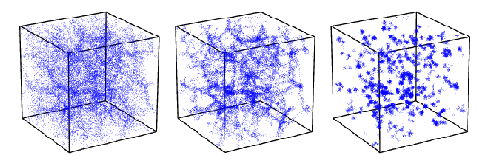}} 
%     \mbox{\hskip -0.0truecm\includegraphics[width=0.98\textwidth]{Rien_Weygaert_cosmoalphabetti_LNCS_Fig5.ps}}
     \vskip -0.0truecm
     \caption{\small Three different Voronoi element models, shown in a cubic setting.
       From left to right:  a wall-dominated Voronoi universe, a filamentary Voronoi universe,
       and a cluster-dominated Voronoi universe.}
  \label{fig:vorelm3}
\end{figure*}

The Voronoi clustering models identify the geometric elements of a $3$-dimensional
Voronoi tessellations with the morphological component of the Cosmic Web.
To describe this relationship, we will adhere to the terminology listed in
Table~\ref{tab:vorcomp}, which also lists  the terminology to the corresponding
types of holes in the alpha shape of the particle distribution\footnote{For consistency and clarity, throughout 
this paper we adopt the following nomenclature for the spatial components of the Cosmic Web, of Voronoi tessellations and 
of Delaunay tessellations. The cosmic web consists of {\it voids, walls, filaments, and clusters}. Voronoi tessellations 
consist of {\it cells, faces, lines, and nodes}. Finally, Delaunay tessellations (and alpha shapes) involve {\it tetrahedra, 
triangles, edges, and vertices.}}.

%%%%%%%%%%%%%%%%%%%%%%%%%%%%%%%%%%%%%%%%%%%%%%%%%%%%%%%%%%%%%%%%%%%%%%%%%%%%%%%%%%%%%%
\subsection{Voronoi Element Models}
\label{sec:vorelm}
%%%%%%%%%%%%%%%%%%%%%%%%%%%%%%%%%%%%%%%%%%%%%%%%%%%%%%%%%%%%%%%%%%%%%%%%%%%%%%%%%%%%%%

These are fully heuristic, user-specified spatial galaxy distributions
within the cells, faces, lines, and nodes of a Voronoi tessellation.
They are obtained by projecting the initially randomly distributed $N$ model galaxies
onto the relevant Voronoi face, line, or node.
We can also retain a galaxy within the Voronoi cell in which it is located.
The Voronoi element models are particularly apt for studying systematic properties of spatial 
galaxy distributions confined to one or more structural elements of nontrivial
geometric spatial patterns. 

\emph{Pure} Voronoi element models place their galaxies exclusively inside or near
either faces, lines, or nodes.
In contrast, \emph{mixed} models allow combinations in which distributions
inside or near cells, faces, lines, and nodes are superimposed.
These include particles located in four distinct structural components:
\begin{itemize}
  \item  \emph{field} particles located in the interior of Voronoi cells,
  \item  \emph{wall} particles within and around the Voronoi faces,
  \item  \emph{filament} particles within and around the Voronoi lines,
  \item  \emph{cluster} particles within and around the Voronoi nodes.
\end{itemize}
The characteristics of the spatial distributions in the mixed models can be 
varied and tuned according to desired fractions of galaxies of each type.
These fractions are free parameters that can be specified by the user;
see Figure~\ref{fig:vorelm3}.

\begin{table}[h]
%\caption{Voronoi elements and Cosmic Web components}
\begin{center}
\begin{tabular}{||l|l||l||}
\hline 
%\hline
&&\\
\hskip 0.5truecm Cosmic Web  \hskip 1.0truecm & \hskip 0.25truecm Voronoi \hskip 2.0truecm &  \hskip 0.25truecm Alpha \hskip 2.0truecm\\
%\hskip 0.5truecm Component \hskip 1.0truecm & \hskip 0.25truecm Element \hskip 2.0truecm &  \hskip 0.25truecm \\
&&\\
\hline 
&&\\
\hskip 0.25truecm voids, field \hskip 1.0truecm & \hskip 0.25truecm cell \hskip 2.0truecm & \hskip 0.25truecm void \\
&&\\
\hskip 0.25truecm walls, sheets, superclusters \hskip 1.0truecm & \hskip 0.25truecm face \hskip 2.0truecm &  \hskip 0.25truecm tunnel\\
&&\\
\hskip 0.25truecm filaments, superclusters \hskip 1.0truecm & \hskip 0.25truecm line \hskip 2.0truecm & \hskip 0.25truecm gap \\
&&\\
\hskip 0.25truecm clusters \hskip 1.0truecm & \hskip 0.25truecm node \hskip 2.0truecm &  \\
&&\\
%\hline
\hline
\end{tabular}
\end{center}
\caption{Identification of morphological components of the Cosmic Web (lefthand column) with geometric Voronoi tessellation elements 
(central column). The righthand column list the terminology for the corresponding types of holes in the alpha shape of the 
particle distribution.}
\vskip -0.5truecm
\label{tab:vorcomp}
\end{table}
%\begin{table}[h]
% \begin{center}
%  \begin{tabular}{l|l|l}
%    Cosmic Web     &  Voronoi &  Alpha      \\ \hline 
%    field          &  cell    &  void       \\
%    wall           &  face    &  tunnel     \\
%    filament       &  line    &  gap        \\
%    cluster        &  node    &       
%  \end{tabular}
% \end{center}
% \caption{Terminology for the components of the Cosmic Web,
%   the Voronoi tessellation of a set of nuclei,
%   and the holes in the alpha shape of the particle/galaxy distribution.}
% \label{tab:vorcomp}
%\end{table}
%%%%%%%%%%%%%%%%%%%%%%%%%%%%%%%%%%%%%%%%%%%%%%%%%%%%%%%%%%%%%%%%%%%%%%%%%%%%%%%%%%%%%%
\subsection{Voronoi Evolution Models}
\label{sec:vorkinm}
%%%%%%%%%%%%%%%%%%%%%%%%%%%%%%%%%%%%%%%%%%%%%%%%%%%%%%%%%%%%%%%%%%%%%%%%%%%%%%%%%%%%%%

The second class we consider are the \emph{Voronoi evolution models}.
They provide web-like galaxy distributions mimicking the outcome of realistic
cosmic structure formation scenarios.
They are based upon the notion that voids play a key organizational role in the development 
of structure, causing the universe to resemble a soapsud of expanding bubbles \cite{icke1984}. 
While the galaxies move away from the void centers, and stream towards the walls,
filaments, and clusters, the fractions of galaxies in or near the cells, faces, lines, and nodes
evolve continuously.
The details of the model realization depends on the specified time evolution.

Within the class of Voronoi evolution models, the most representative and most frequently
used are the \emph{Voronoi kinematic models}.
Forming the idealized and asymptotic description of the outcome of a
hierarchical gravitational structure formation process,
they simulate the asymptotic web-like galaxy distribution implied by the hierarchical
void formation process by assuming a single-size dominated void population. 
Within a void, the mean distance between galaxies increases with time.
Before a galaxy enters an adjacent cell, the velocity component perpendicular
to the otherwise crossed face disappears.
Thereafter, the galaxy continues to move within the face.
Before it enters the next cell, the velocity component perpendicular to the
otherwise crossed edge disappears.
The galaxy continues along a filament and, finally, comes to rest at a node.
The resulting evolutionary progression within the Voronoi kinematic model proceeds 
from an almost featureless random distribution towards a distribution in which matter 
ultimately aggregates into conspicuous compact cluster-like clumps.

The steadily increasing contrast of the various structural features is accompanied
by a gradual shift in the topology of the distribution.
The virtually uniform and featureless particle distribution at the beginning
ultimately unfolds into a highly clumped distribution of clusters.
This evolution involves a gradual progression via a wall-like through a 
filamentary towards an ultimate cluster-dominated matter distribution.

\begin{figure*}[h]
 \begin{center}
      \mbox{\hskip -0.5truecm\includegraphics[bb=0 0 436 372,width=0.99\textwidth]{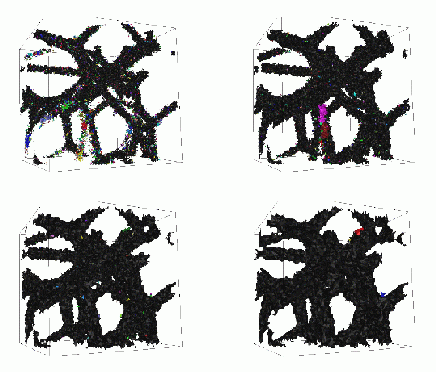}} 
%  \mbox{\hskip -0.5truecm\includegraphics[width=0.99\textwidth]{Rien_Weygaert_cosmoalphabetti_LNCS_Fig6.ps}}
  \vskip -0.0truecm
  \caption{Four alpha shapes of a Voronoi filament model consisting of $200,000$ particles
    in a periodic box of size $50 \Mpch$ with $8$ Voronoi cells.
    We use colors to highlight different components.
    From top left to bottom right: $10^4 \cdot \alpha = 0.5, 1.0, 2.0, 4.0$.}
  \label{fig:vorfilalpha}
 \end{center}
 \vskip -1.0truecm
\end{figure*}
%%%%%%%%%%%%%%%%%%%%%%%%%%%%%%%%%%%%%%%%%%%%%%%%%%%%%%%%%%%%%%%%%%%%%%%%%%%%%%%%%%
%%%%%%%%%%%%%%%%%%%%%%%%%%%%%%%%%%%%%%%%%%%%%%%%%%%%%%%%%%%%%%%%%%%%%%%%%%%%%%%%%%
\medskip
\section{Topological Analysis of Voronoi Universes}
\label{sec:voronoibetti}
%%%%%%%%%%%%%%%%%%%%%%%%%%%%%%%%%%%%%%%%%%%%%%%%%%%%%%%%%%%%%%%%%%%%%%%%%%%%%%%%%%
%%%%%%%%%%%%%%%%%%%%%%%%%%%%%%%%%%%%%%%%%%%%%%%%%%%%%%%%%%%%%%%%%%%%%%%%%%%%%%%%%%

In this section, we study the systematic behavior of the Betti numbers of the
alpha shapes of Voronoi clustering models; see \cite{eldering2006,vegter2010}.
For each point sample, we investigate the alpha shape for the full range of the $\alpha$ parameter.
We generate six Voronoi clustering models, each consisting of $200,000$ particles 
within a periodic box of size $50 \Mpch$.
Of each model, we make two realizations, one with $8$ and the other with $64$ nuclei.
We start with two pure Voronoi element models:  a wall model, and a filament model. 
In addition, we study four Voronoi kinematic models, ranging from a mildly evolved to a strongly 
evolved configuration.
In each case, the clusters, filaments, and walls have a finite Gaussian width of $1.0 \Mpch$. 

An impression of the sequence of alpha shapes may be gained from the four panels in Figure~\ref{fig:vorfilalpha}.
For the smallest value of $\alpha$, we see that the simplices
delineate nearly all the filaments in the particle distribution.
As $\alpha$ increases, going from the top-left panel down to the bottom right panel,
we find that the alpha shape fills in the walls.
For even larger values of $\alpha$, the alpha shape includes the large Delaunay simplices
that cover the interior of the Voronoi cells.
It is a beautiful illustration of the way in which alpha shapes define, 
as it were, naturally evolving surfaces that are sensitive to every detail of the 
morphological structure of the cosmic matter distribution. 

\begin{figure*}
 \vskip -0.5truecm
 \begin{center}
  \mbox{\hskip -0.4truecm\includegraphics[bb=0 0 1916 672,width=1.05\textwidth]{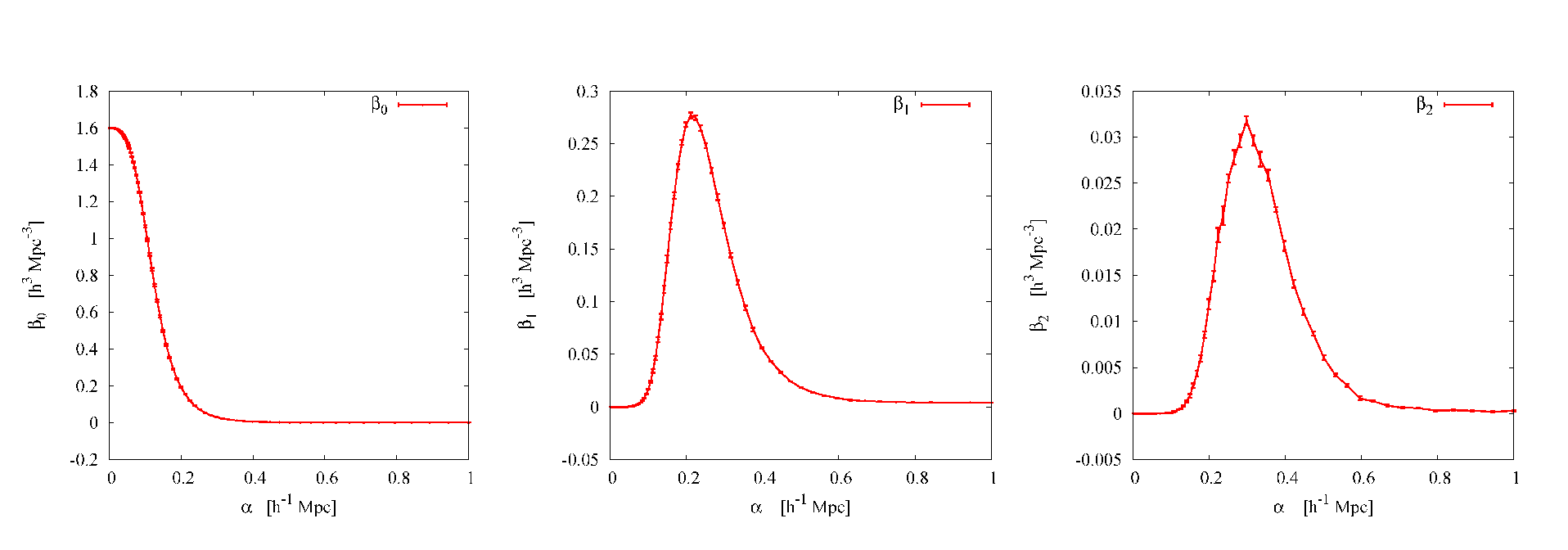}}
  \vskip -0.0truecm
  \caption{Topological analysis of the Voronoi filament model illustrated by its
    alpha shape shown in the upper left panel of Figure \protect{\ref{fig:vorfilalpha}}.
    From left to right:  $\Betti_0$, $\Betti_1$, and $\Betti_2$ counting the
    components, tunnels, and voids of the alpha shape. The realization contains 
    $64$ nuclei or cells.} 
  \label{fig:vorfiltop}
 \end{center}
 \vskip -1.0truecm
\end{figure*}

%%%%%%%%%%%%%%%%%%%%%%%%%%%%%%%%%%%%%%%%%%%%%%%%%%%%%%%%%%%%%%%%%%%%%%%%%%%%%%%%%%%%%%
\subsection{Filament Model Topology}
%%%%%%%%%%%%%%%%%%%%%%%%%%%%%%%%%%%%%%%%%%%%%%%%%%%%%%%%%%%%%%%%%%%%%%%%%%%%%%%%%%%%%%

We take the Voronoi filament model as a case study,
investigating its topology by following the behavior of the three Betti numbers
as functions of the parameter $\alpha$.

Figure~\ref{fig:vorfiltop} shows the relation between the Betti numbers
of the alpha shape and the value of $\alpha$. 
The zeroth Betti number, $\Betti_0$, counts the components.
Equivalently, $\Betti_0 - 1$ counts the gaps between the components.
In the current context, the latter interpretation is preferred as we focus on the holes
left by the alpha shape.
Starting with $\Betti_0 = 200,000$ at $\alpha = 0$, the zeroth Betti number
gradually decreases to $1$ as the components merge into progressively larger entities.

The first Betti number, $\Betti_1$, counts the tunnels. 
At first, it increases steeply when edges are added to the alpha complex,
some of which bridge gaps while others form tunnels.
After $\Betti_1$ reaches its maximum for $\alpha$ roughly equal to
$0.25$ times $10^{-4}$, the number of tunnels decreases sharply
as triangles enter the alpha complex in large numbers.

The second Betti number, $\Betti_2$, counts the voids in the alpha shape.
In the case of the Voronoi filament model, its behavior resembles that of $\Betti_1$:
a peaked distribution around a moderate value of $\alpha$.
It increases when the entering triangles are done closing tunnels and start creating voids.
However, when $\alpha$ is large enough to add tetrahedra,
these voids start to fill up and $\Betti_2$ decreases again, reaching zero eventually.
Notice that in the given example of the Voronoi filament model,
$\Betti_2$ reaches its maximum for $\alpha$ roughly equal to
$0.45$ times $10^{-4}$, which lies substantially beyond the peak in the $\Betti_1$ distribution;
see also Figure~\ref{fig:vorclustbeta2}.

%%%%%%%%%%%%%%%%%%%%%%%%%%%%%%%%%%%%%%%%%%%%%%%%%%%%%%%%%%%%%%%%%%%%%%%%%%%%%%%%%%%%%%
\subsection{Void Evolution}
%%%%%%%%%%%%%%%%%%%%%%%%%%%%%%%%%%%%%%%%%%%%%%%%%%%%%%%%%%%%%%%%%%%%%%%%%%%%%%%%%%%%%%

Having assessed one particular Voronoi clustering model in detail, we turn our attention
to the differences between the models.
While this is still the subject of ongoing research, we find substantial differences
on a few particular aspects.
For example, $\Betti_0$ decreases monotonically with $\alpha$ for all models,
but the range over which the number of gaps is substantially positive,
and the rate with which it decrease are highly sensitive to the underlying distribution. 
In fact, the approximate derivative, $\partial \Betti_0/\partial \alpha$, 
contains interesting features.
Examples are a minimum and a varying width, both  
potentially interesting for discriminating between the underlying topologies.

\begin{figure*}[h]
 \begin{center}
  \includegraphics[bb=0 0 1514 1121,width=0.98\textwidth]{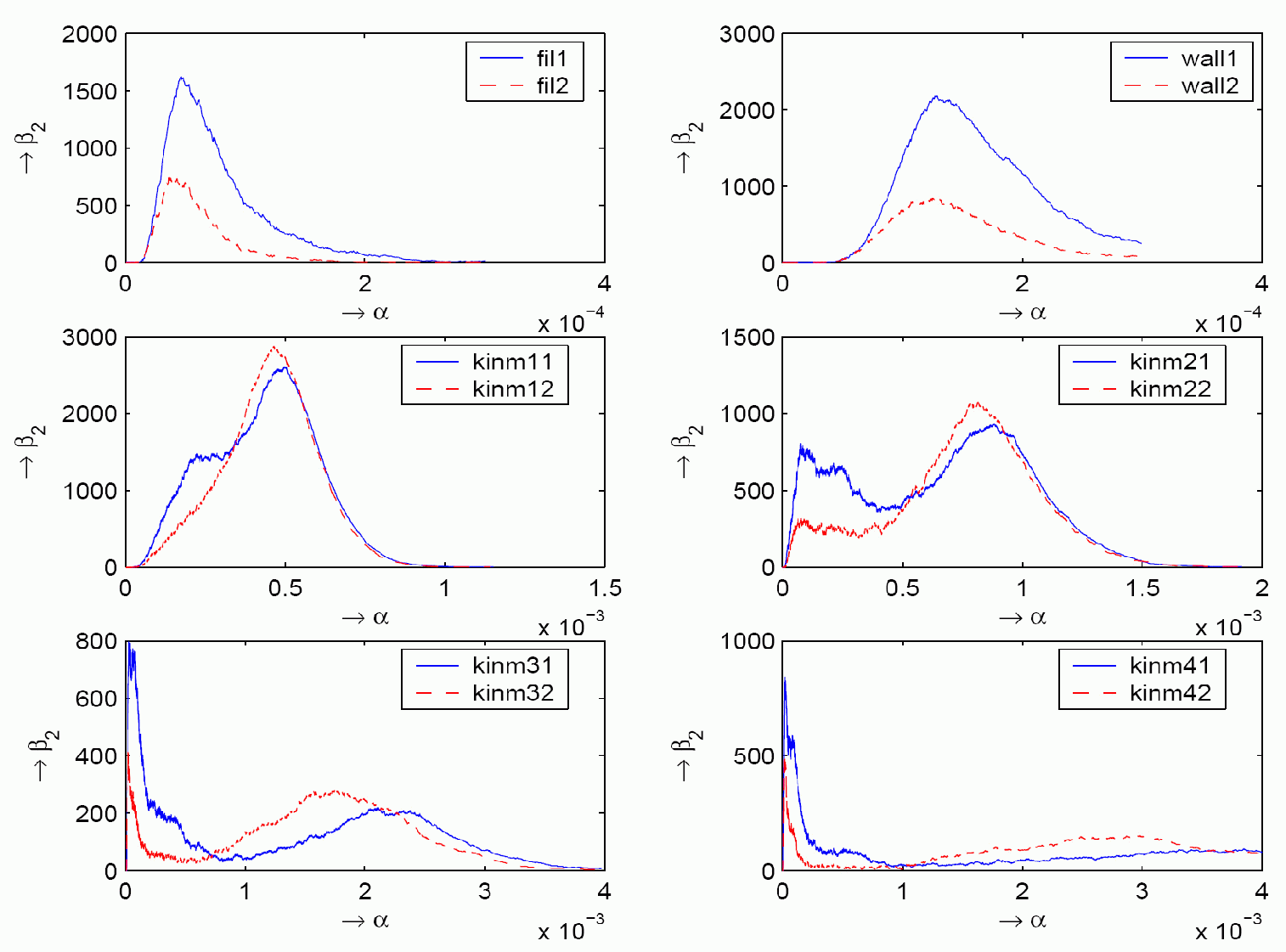} 
  \vskip -0.0truecm
  \caption{The dependence of the second Betti number, $\Betti_2$, on $\alpha$,
    for six different Voronoi clustering models.
    Top left: Voronoi filament model.  Top right: Voronoi wall model.
    Centre left to bottom right: four stages of the Voronoi kinematic model,
    going from a moderately evolved model dominated by walls (center left) to a 
    highly evolved model dominated by filaments and clusters (bottom right).
    Blue lines: realizations with $8$ nuclei or cells.
    Red lines: realizations with $64$ nuclei or cells.}
  \label{fig:vorclustbeta2}
 \end{center}
\end{figure*}

Most interesting is the difference in behavior of the second Betti number.
As one may infer from Figure~\ref{fig:vorclustbeta2}, substantial differences between 
models can be observed.  This concerns the values and range over which 
$\Betti_2$ reaches maxima, as well as new systematic behavior.
For kinematic models, we find two or more peaks, each corresponding
to different morphological components of the particle distribution.
It is revealing to follow the changes in the dependence of $\Betti_2$ on $\alpha$,
as we look at different evolutionary stages. 
The panels from center left to bottom right in Figure~\ref{fig:vorclustbeta2} correspond to 
four different stages of evolution.
The center left panel shows the dependence of $\Betti_2$ on $\alpha$ for
a moderately evolved matter distribution, which is dominated by walls.
The center right panel shows the dependence for a stage at which walls and 
filaments are approximately equally prominent.
In the bottom left panel, the filaments represent more than $40\%$ of the mass, while the walls
and the gradually more prominent clusters each represent around $25\%$ of the particles.
The final, bottom right panel shows the dependence for a highly evolved mass distribution,
with clusters and filaments each representing around $40\%$ of the particles. 

The different morphological patterns of the Voronoi kinematic models are reflected 
in the behavior of $\Betti_2$.  In the center left panel, we find a strong peak at 
$\alpha$ roughly equal to $5$ times $10^{-4}$, with a shoulder at lower values.
The peak reflects the voids inside the walls in the distribution,
while the shoulder finds its origin in the somewhat smaller voids inside the filaments.
The identity of the peaks becomes more clear when we turn to the two-peak distribution 
in the center right panel.
The strong peak at $\alpha$ roughly equal to $10^{-4}$ is a manifestation
of the strongly emerged filaments in the matter distribution.
As the shift to filaments 
and clusters continues, we see the rise of a third peak at a much smaller value of 
$\alpha$; see the bottom two panels.
This clearly corresponds to the voids in the high density cluster regions. 

\begin{figure*}[h]
 \vskip -1.0truecm
 \begin{center}
  \includegraphics[bb=0 0 332 1008,angle=-90.0,width=0.98\textwidth]{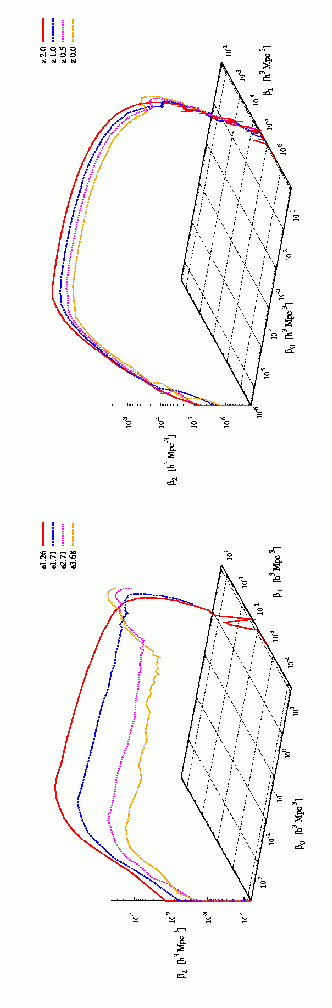} 
% \mbox{\hskip -0.5truecm\includegraphics[angle=-90.0,width=1.05\textwidth]{Rien_Weygaert_cosmoalphabetti_LNCS_Fig9.eps}} 
  \vskip -0.0truecm
  \caption{Left:  the alpha tracks for four stages of the kinematic Voronoi clustering model, 
    evolving from almost uniform (red) to a distribution in which most matter resides in cluster nodes (orange).
    The wall-like (blue) and filamentary (magenta) distributions form intermediate cases.
    Right:  the alpha tracks at different cosmic epochs in an evolving LCDM mass distribution.
    The redshift, $z$, is indicated in the top right corner, with the track at an early epoch
    ($a = 0.17$, $z = 5.0$) bearing resemblance to the early Voronoi kinematic model,
    and the track at the current epoch ($a = 1.0$, $z = 0.0$) resembling the Voronoi kinematic model
    at an intermediate distribution.}
  \label{fig:alphatrack}
 \end{center}
 \vskip -0.0truecm
\end{figure*}
%%%%%%%%%%%%%%%%%%%%%%%%%%%%%%%%%%%%%%%%%%%%%%%%%%%%%%%%%%%%%%%%%%%%%%%%%%%%%%%%%%%%%%
\subsection{Alpha Tracks}
\label{sec:alphatrack}
%%%%%%%%%%%%%%%%%%%%%%%%%%%%%%%%%%%%%%%%%%%%%%%%%%%%%%%%%%%%%%%%%%%%%%%%%%%%%%%%%%%%%%

In Figure~\ref{fig:alphatrack}, we synthesize the homology information
by tracing curves in the $3$-dimensional \emph{Betti space},
whose coordinates are the three Betti numbers.
Each curve is parametrized by $\alpha$, with points
$(\Betti_0 (\alpha), \Betti_1 (\alpha), \Betti_2 (\alpha))$ characterizing
the homology of the alpha shape at this given scale.
We call each curve an \emph{alpha track}, using it to visualize the
combined evolution of the Betti numbers for a given Voronoi cluster model.
Note that the evolution of the Euler characteristic, $\Euler (\alpha)$,
is the projection of an alpha track onto the normal line of the plane 
$\Betti_0 - \Betti_1 + \Betti_2  =  0$.
The richer structure of the full alpha track illustrates why patterns with similar
genus may still have a substantially different topology. 

The left frame of Figure~\ref{fig:alphatrack} shows the alpha tracks for the
four Voronoi kinematic models.
We notice that $\Betti_0$ varies over a substantially larger range of values than $\Betti_1$ and $\Betti_2$.
The tracks reveal that the few components left after an extensive process of merging
still contain a substantial number of tunnels and voids.
More generally, we see that the Betti numbers diminish to small values in sequence:
first $\Betti_0$, then $\Betti_1$, and finally $\Betti_2$.

As the matter distribution evolves, the tracks shift in position. 
The four tracks in the left frame of Figure~\ref{fig:alphatrack} form a good illustration of this effect.
The red dashed track corresponds to an early phase of the kinematic model,
at which the particle distribution is still almost uniform, 
while the orange track represents the final time-step marked by a pattern in 
which nearly all particles reside in the clusters.
In the intermediate stages, most of the particles are located in the walls (blue)
and in the filaments (magenta).
The decreasing number of tunnels and voids at intermediate scales is a manifestation
of the hierarchical formation process, in which 
small scale structures merge into ever larger ones; see e.g.\ \cite{shethwey04}. 

%% THE FOLLOWING PARAGRAPH CONFUSES ME BECAUSE WE SHOW THE TRACK ONLY FOR THE 
%% VORONOI KINEMATIC MODELS ... HOW COME WE CAN COMPARE?
%% An analysis of the evolving Voronoi kinematic models yields alpha tracks 
%% which resemble the plotted tracks for the different Voronoi element models.   
%% As the galaxy distribution in these models evolve from a mostly featureless distribution, 
%% with an alpha track close to that of the red track in Figure~\ref{fig:alphatrack} (left frame), 
%% to that of a predominantly cluster-dominated distribution we see a gradual shift of the alpha 
%% tracks to that resembling the orange track in Figure~\ref{fig:alphatrack} (left frame). 
%% The shift of the alpha tracks is particularly pronounced along the $\Betti_1$ axis: over 
%% nearly the full range of $\alpha$ values the number of tunnels decreases as time proceeds. 
%% We also find a rapid decrease of $\Betti_2$ at intermediate $\alpha$ values, testifying of 
%% a considerable reduction in number of voids.

%%%%%%%%%%%%%%%%%%%%%%%%%%%%%%%%%%%%%%%%%%%%%%%%%%%%%%%%%%%%%%%%%%%%%%%%%%%%%%%%%%
%%%%%%%%%%%%%%%%%%%%%%%%%%%%%%%%%%%%%%%%%%%%%%%%%%%%%%%%%%%%%%%%%%%%%%%%%%%%%%%%%%
\medskip
\section{Topological Analysis of the LCDM Universe}
\label{sec:lcdmbetti}
%%%%%%%%%%%%%%%%%%%%%%%%%%%%%%%%%%%%%%%%%%%%%%%%%%%%%%%%%%%%%%%%%%%%%%%%%%%%%%%%%%
%%%%%%%%%%%%%%%%%%%%%%%%%%%%%%%%%%%%%%%%%%%%%%%%%%%%%%%%%%%%%%%%%%%%%%%%%%%%%%%%%%

Having discussed the scale-dependent Betti numbers in heuristic Voronoi clustering models,
we turn to the analysis of more realistic megaparsec cosmic mass distributions.
These are characterized by an intricate multiscale configuration of anisotropic web-like patterns.
The crucial question is whether we can exploit the topological information toward
determining crucial cosmological parameters, such as the nature of dark energy. 

We concentrate on the analysis of computer simulations of cosmic structure formation in the Universe,
leaving the analysis of the observed distribution in galaxy redshift surveys for the future.
The computer simulations follow the nonlinear evolution of structure as it emerges from the near uniform early Universe.
Once the gravitational clustering process has progressed beyond the initial linear growth phase,
we see the emergence of intricate patterns in the density field; see Figure~\ref{fig:lcdmsimul}.
The near homogeneous initial 
conditions evolve into an increasingly pronounced clustering pattern.

%%%%%%%%%%%%%%%%%%%%%%%%%%%%%%%%%%%%%%%%%%%%%%%%%%%%%%%%%%%%%%%%%%%%%%%%%%%%%%%%%
\subsection{N-body Simulations}
%%%%%%%%%%%%%%%%%%%%%%%%%%%%%%%%%%%%%%%%%%%%%%%%%%%%%%%%%%%%%%%%%%%%%%%%%%%%%%%%%

In cosmological $N$-body simulations, the cosmic mass distribution is represented by a large 
number of particles, which move under the influence of the combined gravitational force of all particles.
The initial conditions (location and velocities of the particles) 
are a realization of the mass distribution expected in the cosmological scenario at hand.
Since the majority of the matter in the Universe is non-dissipative dark matter,
a major aspect of cosmological $N$-body simulations  concerns itself with dark matter particles
that only interact via gravity.  
State-of-the-art computer simulations, such as the Millennium simulation \cite{springmillen2005}, 
count in the order of 10$^{10}$ particles and are run on the most powerful supercomputers 
available to the scientific community. 

Figure~\ref{fig:lcdmsimul} illustrates the matter distribution in the standard, or ``concordance'' 
LCDM cosmological model.
It shows the dark matter distribution in a box of $80 \Mpch$ co-moving size, based on a 
$384^3$ particle $N$-body simulation.
The LCDM cosmological scenario assumes that 
most gravitating matter in the Universe consists of as yet undetected and unidentified cold 
dark matter, accounting for approximately $23\%$ of the energy content of the Universe, i.e.\ $\Omega_{dm}=0.23$.
The normal, baryonic matter, which consists mostly of protons and neutrons,
represents only $4.4\%$ of the energy content of the Universe, i.e.\ $\Omega_b=0.044$. 
Most importantly, the model assumes the presence of a cosmological constant, $\Lambda$,
or equivalent dark energy component, representing $73\%$ of the density of the Universe, 
i.e.\ $\Omega_{\Lambda} = 0.73$; see Section~\ref{sec:debetti}.
The Hubble parameter, which specifies the expansion rate of the Universe,
is taken to be ${\rm H}_0 = 70$ km/s/Mpc.
The amplitude of the initial fluctuation has been normalized to a level at which the current density 
fluctuation on a scale of $8 \Mpch$ is equal to $\sigma_8 = 0.8$. 

The most prominent aspect of the cosmic mass distribution is the intriguing network of 
filaments and high-density clusters, which surround low-density voids.
The mass 
distribution has a distinct multiscale character, marked by clumps over a wide range of 
scales, reflecting the hierarchical evolution of the distribution.

%%%%%%%%%%%%%%%%%%%%%%%%%%%%%%%%%%%%%%%%%%%%%%%%%%%%%%%%%%%%%%%%%%%%%%%%%%%%%%%%%%%%%%
\subsection{Homological Evolution}
%%%%%%%%%%%%%%%%%%%%%%%%%%%%%%%%%%%%%%%%%%%%%%%%%%%%%%%%%%%%%%%%%%%%%%%%%%%%%%%%%%%%%%

\begin{figure*}[h]
 \vskip -0.0truecm
 \begin{center}
  \includegraphics[bb=0 0 684 506,width=0.98\textwidth]{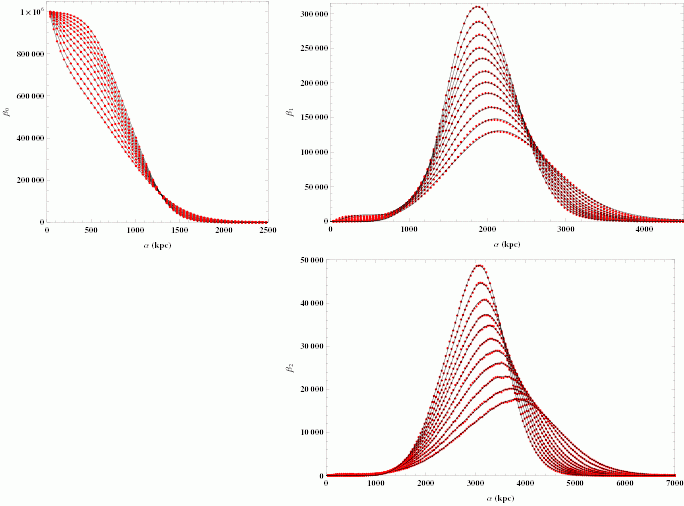} 
%  \mbox{\hskip -0.5truecm\includegraphics[width=1.02\textwidth]{Rien_Weygaert_cosmoalphabetti_LNCS_Fig10.eps}}
  \vskip -0.0truecm
  \caption{The evolution of the Betti numbers in an LCDM universe.
    We show the dependence of $\Betti_p$ on $\alpha$ for dimensions $p = 0, 1, 2$ and
    for eleven expansion factors running from $a = 0.25$ ($z = 3.0$)
    to the current epoch $a = 1.0$ ($z = 0.0$).
    From upper left to lower right:  $\Betti_0(\alpha)$, $\Betti_1(\alpha)$, $\Betti_2(\alpha)$.
    Note that all three curves gradually shift from smaller to larger scales.}
  \label{fig:bettilcdm}
 \end{center}
 \vskip -0.5truecm
\end{figure*}

We follow the developing structure in the LCDM scenario in eleven time-steps
that run from an expansion factor $a = 0.25$ ($z = 3.0$) to $a = 1.0$ ($z = 0.0$),
i.e.\ from around $3.4$ Gigayears after the Big Bang to the current epoch.
At each time-step, we compute the alpha shapes of a randomly sampled subset of the particle distribution 
for $\alpha$ from $0.0$ to $10.0 \Mpch$,
and their Betti numbers, from which we extrapolate the scale-dependent behavior of homology.
The resulting curves of the Betti numbers for dimensions $p = 0, 1, 2$
are shown in Figure~\ref{fig:bettilcdm}.
The curves in the upper left frame show that the value of $\Betti_0$ decreases
for $\alpha < 1.2 \Mpch$, while it increases for larger values of $\alpha$.
This is a manifestation of the hierarchical structure formation process.
It reflects the progressively earlier merging of clumps at small scale
and the progressively later merging of massive components at large scale.
 
The curves of $\Betti_1$ and $\Betti_2$ have a slightly different appearance.
Both start as highly peaked distributions. Most prominent in the near homogeneous 
initial conditions, the peaks represent the imprint of a 
Poisson distribution with a similar particle density. In principle, one should 
remove this imprint via a persistence procedure; see Section~\ref{sec:persist}.
Here we keep to the unfiltered version. As the mass distribution evolves 
under the influence of gravity, the particles get more and more clustered. 
This leads to a decrease in the number of holes on small scales.
%% I DON'T UNDERSTAND THE NEXT SENTENCE.
As these merge into ever larger supercluster complexes, 
their formation goes along with the evacuation of ever larger tunnels and 
voids enclosed by the higher density structures in the web-like network. 

All three curves are marked by a steady shift toward higher values of $\alpha$.
While we find a minor shift of the maximum of $\Betti_1$, the shift is clear for $\Betti_2$.
It is interesting that a similar shift toward larger $\alpha$ values has has been predicted
by Sheth \& van de Weygaert \cite{shethwey04} in the context of their 
dynamical model of the evolving void hierarchy.
In their two-barrier excursion set formalism, this shift is a manifestation of the merging of 
smaller voids into ever larger ones, accompanied by the destruction of a 
large number of small voids inside gravitationally collapsing over-dense regions.

%%%%%%%%%%%%%%%%%%%%%%%%%%%%%%%%%%%%%%%%%%%%%%%%%%%%%%%%%%%%%%%%%%%%%%%%%%%%%%%%%%%%%%
\subsection{Alpha Tracks and Multiscale Homology}
%%%%%%%%%%%%%%%%%%%%%%%%%%%%%%%%%%%%%%%%%%%%%%%%%%%%%%%%%%%%%%%%%%%%%%%%%%%%%%%%%%%%%%

When assessing the corresponding evolution of alpha tracks in Figure~\ref{fig:alphatrack}, 
we find a systematic behavior that is similar to what we have seen for the Voronoi clustering models. 
As the evolution proceeds, we  find fewer tunnels and voids at small scale,
while their numbers are still substantial at values of $\alpha$ at which we have only very few
remaining components.

However, as the mass distribution evolves, the shifts between the alpha tracks 
are relatively small compared to those seen in the Voronoi clustering models. 
The mass distribution retains its multiscale character. The hierarchical 
evolution of the multiscale mass distribution leads to a more or less self-similar 
mapping of the LCDM alpha tracks to higher $\alpha$ values.
At each phase, we find dominant clusters and conspicuous filamentary features,
although their scale shifts as the mass distribution advances from the mildly linear to
the quasi-linear phase. 
This contrasts the evolution of the Voronoi clustering models, which is characterized 
by clear transitions between distinct topological patterns outlined by the mass distribution.

%%%%%%%%%%%%%%%%%%%%%%%%%%%%%%%%%%%%%%%%%%%%%%%%%%%%%%%%%%%%%%%%%%%%%%%%%%%%%%%%%%
%%%%%%%%%%%%%%%%%%%%%%%%%%%%%%%%%%%%%%%%%%%%%%%%%%%%%%%%%%%%%%%%%%%%%%%%%%%%%%%%%%
\medskip
\section{Probing Dark Energy}
\label{sec:debetti}
%%%%%%%%%%%%%%%%%%%%%%%%%%%%%%%%%%%%%%%%%%%%%%%%%%%%%%%%%%%%%%%%%%%%%%%%%%%%%%%%%%
%%%%%%%%%%%%%%%%%%%%%%%%%%%%%%%%%%%%%%%%%%%%%%%%%%%%%%%%%%%%%%%%%%%%%%%%%%%%%%%%%%

To assess the utility of our topological methods, we look at the possibility to use the homology
of the Cosmic Web towards determining the nature of dark energy in the Universe.
The parameters for what might now be called ``the standard model for cosmology''
have been established with remarkable precision. 

However, there remains the great mystery of the nature of the so-called \emph{dark energy},
which appears to make up $73\%$ of the total cosmic energy.
Dark energy, or the cosmological constant, has been dominating the dynamics of the Universe since the 
last 7 Gigayears and has pushed it into an accelerated expansion.  
The simplest model for the dark energy is Einstein's cosmological constant, $\Lambda$: 
it makes a time-independent contribution to the total energy density in the 
Friedman-Lemaitre equations.   Such models are referred to as \emph{LCDM models}.  
However, there are numerous, possibly more plausible models in which the dark energy 
evolves as a function of time.  These models are generally described in terms
of a time- or redshift-dependent function $w(z)$ that describes the history of equation of state of 
the dark energy component, i.e.\ the relation between pressure and density:
\begin{eqnarray}
  p  &=&  w(z) \rho c^2 .
\end{eqnarray}
Different models for dark energy produce different functions $w(z)$,
and a number of simple parametrizations can be found in the literature.
The classical Einstein cosmological constant corresponds to $w(z) = -1$.
\begin{figure}[h]
 \vskip -0.5truecm
 \begin{center}
  \includegraphics[bb=0 0 1310 1121,width=0.98\textwidth]{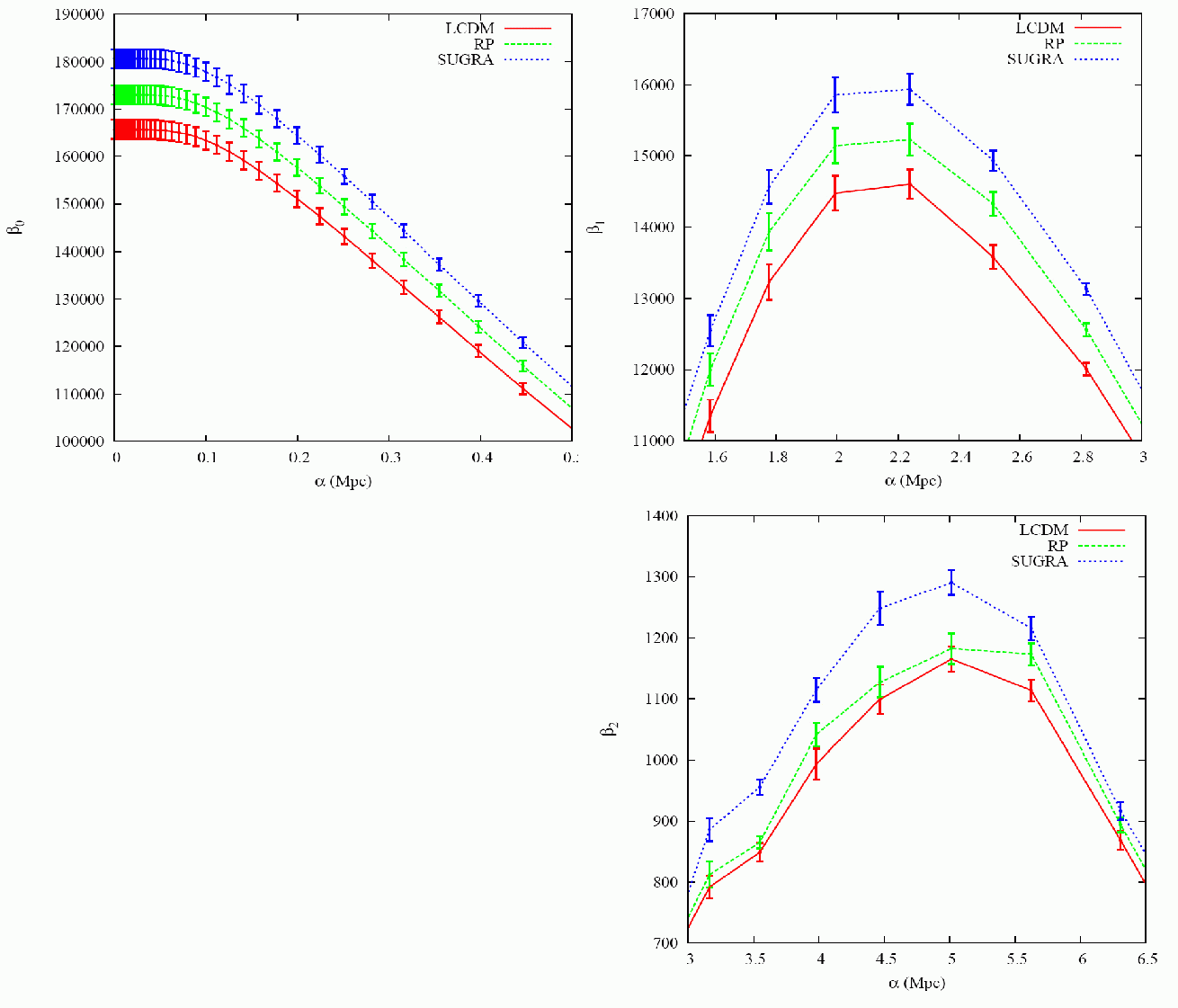} 
  \vskip -0.0truecm
  \caption{Homology and dark energy.
    The Betti number curves for three cosmological models with different dark energy models.
    The curves are determined from the distribution of collapsed dark matter halos in the corresponding scenarios.
    Red: the standard LCDM scenario.  Green: the Ratra-Peebles quintessence model.
    Blue: the supergravity SUGRA model.
    From left to right: $\Betti_0(\alpha)$, $\Betti_1(\alpha)$, and $\Betti_2(\alpha)$.
    The curves demonstrate that homology is capable of using the web-like matter distribution to discriminate 
    between cosmological scenarios with different dark energy content.}
\label{fig:debetti}
\end{center}
\vskip -1.25truecm
\end{figure}
In principle, the Cosmic Web should be one of the strongest differentiators between the various
dark energy models.
As a result of the different growth rates at comparable cosmic epochs, we will see different 
structures in the matter distribution. 
To illustrate this idea, we run three $N$-body simulations with identical initial 
conditions but different dark energy equation of state:
the \emph{standard LCDM model} and two \emph{quintessence models}.
The latter assume that the Universe contains an evolving quintessence scalar field,
whose energy content manifests itself as dark energy.
The two quintessence models are the \emph{Ratra-Peebles (RP) model} and the \emph{SUGRA model};
see \cite{ratra88,deboni11,bos11} for a detailed description.

While the general appearance of the emerging Cosmic Web is structurally similar, 
there are visible differences in the development. On small to medium scales, the quintessence 
models are less clustered; see \cite{weygaert11} for an illustration.
As a result, the alpha shapes contain a larger number of small objects.
In Figure~\ref{fig:debetti}, we see that this manifests itself in values of $\Betti_0$
that are systematically higher for the quintessence RP and SUGRA models than for the LCDM model.
The two right frames in Figure~\ref{fig:debetti} concentrate on the 
medium scale range, $1.5 \Mpch < \alpha < 3.0 \Mpch$ for $\beta_1$ (top righthand panel) and 
$3.0 \Mpch < \alpha < 6.5 \Mpch$ for $\beta_2$ (bottom righthand panel). The error bars on the obtained 
curves are determined on the basis of the obtained values for independent samples.
Particularly encouraging is the fact that homology is sensitive to the subtle differences
we see in the pattern of the Cosmic Web. 
The size of the error bars show that it is feasible to find significant and measurable 
differences between the outcome of the different cosmologies.

%%%%%%%%%%%%%%%%%%%%%%%%%%%%%%%%%%%%%%%%%%%%%%%%%%%%%%%%%%%%%%%%%%%%%%%%%%%%%%%%%%
%%%%%%%%%%%%%%%%%%%%%%%%%%%%%%%%%%%%%%%%%%%%%%%%%%%%%%%%%%%%%%%%%%%%%%%%%%%%%%%%%%
\medskip
\section{Betti Numbers of Gaussian Random Fields}
\label{sec:gaussian}
%%%%%%%%%%%%%%%%%%%%%%%%%%%%%%%%%%%%%%%%%%%%%%%%%%%%%%%%%%%%%%%%%%%%%%%%%%%%%%%%%%
%%%%%%%%%%%%%%%%%%%%%%%%%%%%%%%%%%%%%%%%%%%%%%%%%%%%%%%%%%%%%%%%%%%%%%%%%%%%%%%%%%

To interpret our results for the web-like cosmic matter distribution,
we compare them with those for the initial condition out of which our Universe arose.
The behavior of the Betti numbers in a Gaussian random field
is a reference point for any further assessment of their behavior in
the more complex environment of the Cosmic Web. 

According to the current paradigm, structure in the Universe grew by gravitational
instability out of tiny primordial density perturbations. 
The evidence provided by the temperature fluctuations in the cosmic microwave background 
\cite{smoot1992,bennett2003,spergel2007,komatsu10} suggests that the character of the 
perturbation field is that of a homogeneous and isotropic spatial Gaussian process.
Such primordial Gaussian perturbations in the gravitational potential
are a natural product of an early inflationary phase of our Universe.

Before proceeding to the Betti numbers of a Gaussian random field,
we present the necessary nomenclature, focusing on the three-dimensional situation.
For fundamentals on Gaussian random fields,
we refer to the standard work by Adler and Taylor \cite{adler2007}, 
and for their cosmological application to the seminal study by
Bardeen, Bond, Kaiser, and Szalay \cite{bbks}. 
Here we follow the notation from van de Weygaert and Bertschinger \cite{weyedb96}. 
The first papers on the homology and persistence of Gaussian random fields are
Adler et al.\ \cite{adler2010} and Park et al.\ \cite{park11}; see also \cite{pranav11}.
Here, we discuss some of the main findings, and we refer to the mentioned papers
for more detailed treatments.
We alert the reader to the fact that the analysis in this section concerns itself
with density fields, which is different from earlier sections in which we
studied distance fields defined by particle distributions.

%%%%%%%%%%%%%%%%%%%%%%%%%%%%%%%%%%%%%%%%%%%%%%%%%%%%%%%%%%%%%%%%%%%%%%%%%%%%%%%%%%%%%%
\subsection{Gaussian Random Fields}
%%%%%%%%%%%%%%%%%%%%%%%%%%%%%%%%%%%%%%%%%%%%%%%%%%%%%%%%%%%%%%%%%%%%%%%%%%%%%%%%%%%%%%

A {\it random field}, $f$, on a spatial volume assigns a value, $f({\bf x})$, to each 
location, $\xx$, of that volume. The fields of interest, such as 
the primordial density or velocity field, are smooth and continuous \footnote{In this 
section, the fields $f(\xx)$ may either be the raw unfiltered field or, without loss 
of generality, a filtered field $f_s(\xx)$. A filtered field is a convolution with 
a filter kernel $W(\xx,\yy)$, $f_s(\xx)=\int \diff\yy f(\yy) W(\xx,\yy)$.}. The stochastic 
properties of a random field are defined by its \emph{$N$-point joint probabilities}, 
where $N$ can be any arbitrary positive integer. To denote them, we write 
$\XX=(\xx_1,\xx_2,\cdots,\xx_N)$ for a vector of $N$ points and $\ff = (f_1, f_2, \ldots, f_N)$ for a vector 
of $N$ field values. The joint probability is
\begin{eqnarray}
  \Probability [f(\xx_1)=f_1,\ldots,f(\xx_N)=f_N] &=& \ProbDensity_{\XX}( \ff) \diff \ff\,,\nonumber\\
  \label{eqn:probability}
\end{eqnarray}
%\begin{eqnarray*}
%  \Probability [f(\xx_1)=f_1,\ldots,f(\xx_N)=f_N] &=&  \frac{\exp \left[ -\frac{1}{2} \sum_{i=1}^N
%                                     \sum_{j=1}^N f_i (\Matrix^{-1})_{ij} f_j \right]}
%                        {[(2 \pi)^N (\det \Matrix)]^{1/2}}
%                   \prod_{k=1}^N \diff f_k ,
%\end{eqnarray*}
which is the probability that the field $f$ at the locations $\xx_i$ 
has values in the range $f_i$ to $f_i+\diff f_i$, for each $1 \leq i \leq N$. 
%The measure 
%$\diff \ff$ is the interval 
%\begin{equation}
%\diff \ff\ = \ [f_1,f_1+\diff f_1]\times[f_2,f_2+\diff f_2]\times \cdots \times[f_N,f_N+\diff f_N]\,. 
%\end{equation}
%The probability $\Probability [f(\xx_1)=f_1,\ldots,f(\xx_N)=f_N]$ is the integral of
%$\ProbDensity_{\XX}(\ff): \Rspace^N \to \Rspace$ over the interval $\diff \ff$, 
Here, $\ProbDensity_{\XX}(\ff)$ is the probability density for the field realization 
vector $\ff$ at the location vector $\XX=(\xx_1,\xx_2,\cdots,\xx_N)$. 
For a \emph{Gaussian random field}, the joint probabilities for $N = 1$ and $N = 2$ determine 
all others. Specifically, the probability density functions take the simple form 
\begin{eqnarray}
  \ProbDensity_{\XX}  (\ff)  &=&  C \cdot \exp \left[ {- \ff \Matrix^{-1} \ff^T / 2} \right]\,,
  \label{eqn:distribution}
\end{eqnarray}
where $C =  1 / [(2 \pi)^N (\det \Matrix)]^{1/2}$ normalizes the expression,
making sure that the integral of $\ProbDensity_{\XX} (\ff)$, over all $\ff \in \Rspace^N$, is equal to $1$.
Here, we assume that each $1$-point distribution is Gaussian with zero mean. 
The matrix $\Matrix^{-1}$ is the inverse of the $N \times N$
covariance matrix with entries
\begin{eqnarray}
  M_{ij}  &=&  \langle f(\xx_i) f(\xx_j) \rangle\,,
\end{eqnarray}
in which the angle bracket denotes the ensemble average of the product,
over the 2-point probability density function. In effect, $\Matrix$ is the generalization 
of the variance of a $1$-point normal distribution, and we indeed have 
$\Matrix = [ \sigma_0^2 ]$ for the case $N = 1$.

Equation \eqref{eqn:distribution} shows that a Gaussian random is fully specified by 
the autocorrelation function, $\xi(r)$, which expresses the correlation 
between the density values at two points separated by a distance $r=|\bf r|$,   
\begin{equation}
\xi(r)\,=\,\xi(|{\bf r}|)\,\equiv\,\langle f(\xx) f(\xx+{\bf r}) \rangle \,.
\label{eq:xi}
\end{equation}
Here we use the \emph{statistical cosmological principle}, which states that 
statistical properties of e.g. the cosmic density distribution in the Universe are 
uniform throughout the Universe. It means that the distribution functions and 
moments of fields are the same in each direction and at each location. The latter implies 
that ensemble averages depend only on one parameter, namely the distance between the points. 
In other words, the entries in the matrix are the values of the \emph{autocorrelation function} for the 
distance between the points: 
$M_{ij} = \xi (r_{ij})$, with $r_{ij} = \| {\xx_i - \xx_j} \|\,.$

An impression of a typical Gaussian random field may be obtained from the
three-dimensional realization shown in the left panel of Figure~\ref{fig:gaussbetti}.
The field is chosen using the power spectrum for the standard LCDM cosmology
with some Gaussian filtering; see e.g.\ \cite{bbks,eisenstein98}.
The image illustrates that Gaussian random fields are symmetric, i.e.\ negative values 
are as likely as positive values. 

\subsection{Power Spectrum}
A stochastic random density field is composed of a spectrum of density fluctuations, each of a 
different scale. The relative amplitudes of small-scale and large-scale fluctuations is of decisive 
influence on the outcome of the subsequent gravitational evolution of the density field and on the 
emerging patterns in the spatial density distribution. 

To describe the multiscale composition of a density field, we write it as a sum of individual harmonic waves, 
i.e. in terms of its Fourier sum. Each of the waves is specified by its \emph{wave vector} $\kk=(k_x,k_y,k_z) \in \Rspace^3$, 
describing the direction and spatial frequency of the wave. The latter is determined by the \emph{magnitude} of the wave vector,
\begin{equation}
k\,=\,|\kk|\,=\,\sqrt{k_x^2+k_y^2+k_z^2}\,, 
\end{equation}
which is the inverse of the wavelength, $\lambda=2\pi/k$. 
Subsequently, we write the field $f(\xx)$ as the Fourier integral, 
\begin{eqnarray}
f({\xx})=\int \d3k \ {\hatf}(\kk)\,e^{-\imunit \kk\cdot \xx}\,,
\end{eqnarray}
where $e^{\imunit \varphi} = \cos \varphi + \imunit \sin \varphi$, as usual, and ${\kk}\cdot{\xx}$ the inner product 
between the wave vector, $\kk$, and position vector, $\xx$. The Fourier components, ${\hatf}(\kk) \in \Cspace$, represent 
the contributions by the harmonic wave $\exp({\imunit \kk\cdot\xx})$ to the field $f(\xx)$. They are 
given by the inverse Fourier transform,
\begin{eqnarray}
{\hatf}({\kk})=\int \dx \ f(\xx)\,e^{\imunit \kk\cdot \xx}\,.
\end{eqnarray}
Because the density field is always real, $f(\xx) \in \Rspace$, the Fourier components  
${\hatf}(\kk) \in \Cspace$ obey the symmetry constraint  
\begin{equation}
{\hatf}(\kk)\,=\,{\hatf}^\ast(-\kk)\,. 
\end{equation}
It identifies ${\hatf}(\kk)$ with the complex conjugate of the Fourier component of the wave vector $-\kk$,
i.e.\ of the wave with the same spatial frequency oriented in the opposite direction.

The \emph{power spectrum} is formally defined as the mean square of the Fourier components, $\hatf (\kk)$, of 
the field. It can be computed from the Fourier components using the Dirac delta function, $\delta_D: \Rspace \to \Rspace$, 
which is the limit of the normal distribution, whose variance goes to zero. Most importantly, the integral of the 
Dirac delta function is assumed to be $1$. Now, we have
\begin{eqnarray}
  \langle {\hatf}({\kk}) {\hatf}({\kk}')\rangle\,=\,(2\pi)^{3/2}\,P(k)\,\delta_D({\kk}-{\kk}')\,,
\label{eq:powerpk}
\end{eqnarray}
and we can compute the power spectrum accordingly. We note that because of the statistical isotropy of the field, 
we have $P(\kk) = P(\kk')$, whenever $|\kk| = |\kk'|$.  We can therefore introduce the $1$-dimensional power spectrum, 
defined by $P(k) = P(\kk)$ for every $\kk$, which we refer to by the same name, for convenience.

The \emph{power spectrum} is the Fourier transform of the autocorrelation function $\xi(\xx)$, which is  
straightforward to infer from equations \eqref{eq:xi} and \eqref{eq:powerpk}, 
\begin{eqnarray}
\xi(x)\,=\,\int \d3k \ P(k)  {\rm e}^{-{\imunit}{\kk}\cdot{\xx}}\,,
\end{eqnarray}
where $\xx$ is a point in the volume with $|\xx| = x$.

Because of the (statistical) isotropy of the field $f(\xx)$, it is straightforward to perform the 
angular part of the integral over the 2-sphere, to yield the following 
integral expression for the autocorrelation function:
\begin{equation}
\xi(x)\ = \ \int {\displaystyle k^2\,\diff k \over \displaystyle 2\pi^2}\,P(k)\,{\displaystyle \sin{kx} \over \displaystyle kx}\,,
\end{equation}
with $k=|\kk|$ and $x=\xx$.. 

From the above we find that the power spectrum is a complete characterization of a homogeneous and isotropic Gaussian random field. 
Within its cosmological context, the power spectrum encapsulates a wealth of information on the parameters and content of the Universe.
Its measurement is therefore considered to be a central key for the understanding of the origin of the cosmos. 

%%%%%%%%%%%%%%%%%%%%%%%%%%%%%%%%%%%%%%%%%%%%%%%%%%%%%%%%%%%%%%%%%%%%%%%%%%%%%%%%%%%%%%
\subsection{Fourier Decomposition}
%%%%%%%%%%%%%%%%%%%%%%%%%%%%%%%%%%%%%%%%%%%%%%%%%%%%%%%%%%%%%%%%%%%%%%%%%%%%%%%%%%%%%%

The unique nature of Gaussian fluctuations is particularly apparent when considering
the stochastic distribution of the field in terms of its Fourier decomposition. The 
assumption of Gaussianity means that the stochastic distribution of the Fourier components 
\begin{eqnarray}
{\hatf}(\kk) &\,=\,& \hatf_{\real}(\kk)+\imunit \hatf_{\imaginary}(\kk)\nonumber\\
             &\,=\,& |\hatf(\kk)|\,e^{\imunit\,\theta(\kk)}
\end{eqnarray}
involves a random phase, $\theta (\kk)$, in the complex plane.  This follows from the mutual 
independence of its real and imaginary parts, each of which is a Gaussian variable. 
While the phase, $\theta (\kk)$, has a uniform distribution over the interval $[0,2\pi]$, the amplitude, 
$r = |\hatf (\kk)|$, has a Rayleigh distribution:
\begin{eqnarray}
  \Prob (r) &=&  \frac{r}{s} \exp{\left[-\frac{r^2}{2s} \right]} ,
\end{eqnarray}
where $s = P(k)$ is the power spectrum value at spatial frequency $k$. 

Subsequently determining the probability of the field realization $f(\xx)$ 
in terms of the probability density function of its Fourier decomposition,
one finds the interesting result that it is the product of the individual probability 
distributions of each of the individual Fourier components $\hatf (\kk)$, 
each Gaussian distributed with zero mean and variance $\sigma^2 (k) = P(k)$. 
It is most straightforward to appreciate this by assessing the probability 
of the entire field $f(\xx)$. 

To infer the probability $\ProbDensity[f]$ of $f(\xx)$, we take the $N$-point joint 
probabilities for the limit of $N \rightarrow \infty$ with uniform spatial sampling, 
the summations appearing in equation \eqref{eqn:distribution} may be turned into 
integrals. The resulting expression for the infinitesimal probability 
$\ProbDensity[f]$ with measure $\DMeasure[f]$, 
\begin{equation}
\ProbDensity[f]\,=\,{\rm e}^{-S[f]}\,\DMeasure[f]\,,
\end{equation}
\noindent involves the probability density of the field, $\exp\left(-S[f]\right)$.
The square brackets in $\ProbDensity[f]$ and $S[f]$ indicate that these are functionals, 
i.e., they map the complete function $f({\xx})$ to one number. The probability density 
is similar to the quantum-mechanical partition function in path integral form, where $S$ 
is the action functional (see \cite{weyedb96}). For a Gaussian random field the expression for 
the action $S$ may be inferred from equation \eqref{eqn:distribution}, 
\begin{equation}
S[f]\,=\,{\displaystyle {1 \over 2}}\,\int \diff\xx_1 \int\diff \xx_2\, f(\xx_1) K(\xx_1-\xx_2) f(\xx_2)\,,
\end{equation}
\noindent where $K$ is the functional inverse of the correlation function $\xi$, 
\begin{equation}
\int \diff \xx\,K(\xx_1-\xx)\xi(\xx-\xx_2)\,=\,\delta_D(\xx_1-\xx_2)\,,
\end{equation}
\noindent and $\delta_{\rm D}$ the Dirac delta function. By transforming this 
expression for the action $S[f]$ to Fourier space, one finds the integral 
expression (see \cite{weyedb96}, appendix B),
\begin{equation}
S[f]\,=\,\int \d3k\,{\displaystyle |{\hat f}({\kk})|^2 \over \displaystyle 2 P(k)}\,.
\end{equation}
%\begin{eqnarray}
%  \Prob_{\XX}  &\propto&  \exp{\left[ - \sum_i \frac{ r_i^2 }{ 2 s_i } \right]} \nonumber\\
%           &\propto&  \prod_i \exp{\left[ - \frac{ r_i^2 }{ 2 s_i } \right]} ,
%\end{eqnarray}
%where $r_i = |\hatf(\kk_i)|$ and $s_i = P(k_i)$; see \cite{weyedb96}. 
This immediately demonstrates that a Gaussian field has the unique property of its 
Fourier components being mutually independent. It has the practical virtue of considerably 
simplifying the construction of Gaussian fields by sampling its Fourier components.

\begin{figure*}[h]
\begin{center}
\includegraphics[bb=0 0 1355 622,width=0.99\textwidth]{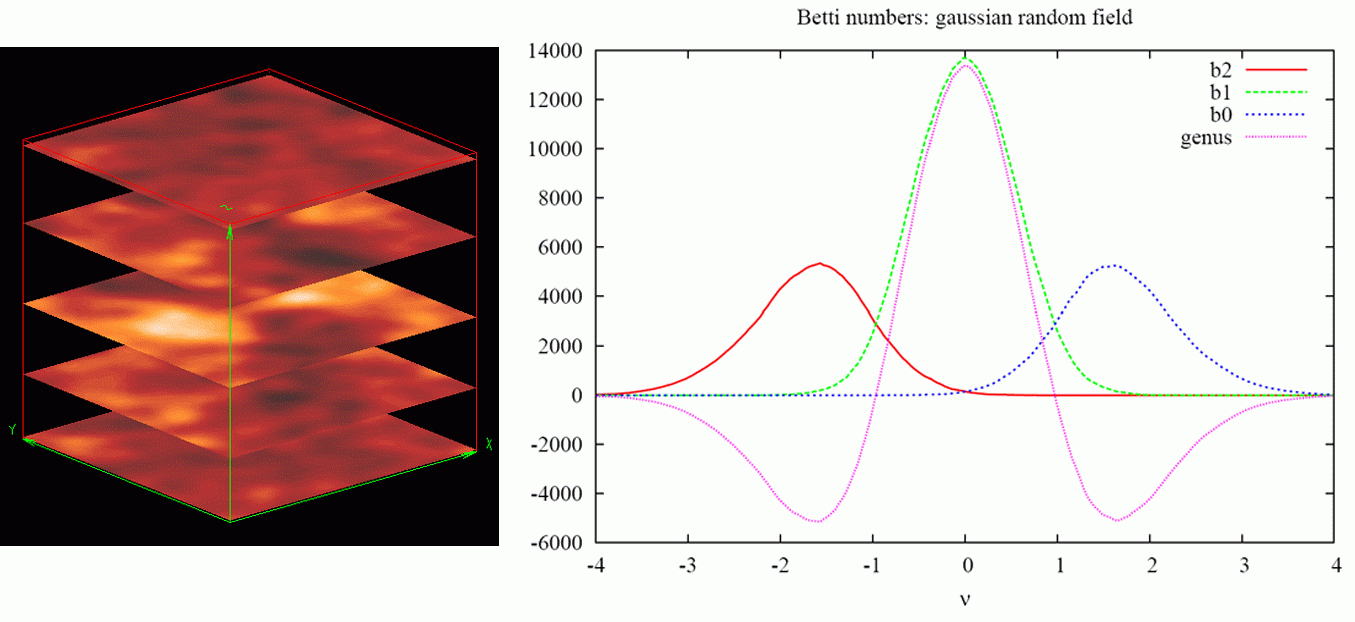} 
\vskip -0.0truecm
\caption{Left: five slices through a realization of a Gaussian random field,
    with the color reflecting the density, running from bright yellow (high) to dark red (low). 
    It is an LCDM density field in a box of $100 \Mpch$, Gaussian filtered on a scale of $5 \Mpch$.
    Right: expected Betti numbers and expected \emph{reduced genus} of the superlevel sets in a 
    Gaussian random field, as functions of the density threshold, $\nu$.}
\label{fig:gaussbetti}
\end{center}
\vskip -1.0truecm
\end{figure*}

%%%%%%%%%%%%%%%%%%%%%%%%%%%%%%%%%%%%%%%%%%%%%%%%%%%%%%%%%%%%%%%%%%%%%%%%%%%%%%%%%%%%%%
\subsection{Betti Numbers}
%%%%%%%%%%%%%%%%%%%%%%%%%%%%%%%%%%%%%%%%%%%%%%%%%%%%%%%%%%%%%%%%%%%%%%%%%%%%%%%%%%%%%%

To determine the Betti numbers of a Gaussian field, we consider the density values
sampled on a regular grid and assess the topology of the superlevel sets. 
We adopt a $64^3$ grid, and sample the field inside a box of size $100 \Mpch$. 
The density values are smoothed on a filter scale $2.0 \Mpch$.
Subsequently, we determine the Betti numbers of superlevel sets,
the agglomerate of regions consisting of the voxels whose density is in excess
of $\nu \sigma(R_f)$, where $\sigma^2 (R_f)$ is the variance of the density field
on the filter scale $R_f$. 

Having outlined the superlevel sets, parameterized by the threshold value $\nu$, we 
subsequently determine their Betti numbers. To this end, we use the code of Kerber \cite{bendich10}, 
which determines the cycles for the superlevel sets sampled on a grid, and 
subsequently produces the corresponding homology groups and their ranks, i.e.\ the 
Betti numbers $\Betti_p(\nu)$. By following this procedure over a threshold level 
range $\nu\in[-5.0,5.0]$, we can evaluate the systematic behavior of the Betti numbers 
as function of $\nu$. Note that a different algorithm was used in \cite{park11}. The 
fact that the results are similar is reassuring. 

The right panel of Figure~\ref{fig:gaussbetti} depicts the curves for $\Betti_0(\nu)$, 
$\Betti_1(\nu)$, and $\Betti_2(\nu)$.
The first major impression is the dominant peak of $\Betti_1(\nu)$ at $\nu=0$
(green dashed curve). This relates to the fact that tunnels prevail 
the topological structure of the outlined regions, which in a Gaussian field are known 
to possess an intricate sponge-like topology \cite{gott1986}.
The region of points with density value at most $\nu = 0$ consists of very few
massive structures that percolate through the volume and, by symmetry, the same
is true for the region of points with density value at least $\nu = 0$.
It is not hard to imagine that this goes along with a large number of tunnels. 

Going away from $\nu = 0$ in either direction, we see that $\Betti_1$ quickly drops to a negligible level.
This marks the transition towards a topology marked by individual objects, either clumps for positive 
$\nu$ or voids for negative $\nu$, and this goes along with the disappearance of tunnels. 
Proceeding to higher levels, we find a quick rise of $\Betti_0$ (blue dotted curve), 
the number of over-dense regions in the volume.
This is a manifestation of the breaking up into individual high-density regions of
the percolating volume at $\nu = 0$.
The number of components reaches its maximum for $\nu$ approximately equal to $\sqrt{3}$.
Beyond that value, the number of individual clumps decreases rapidly, 
as a result of the decreasing probability of regions having a density that high.
It is interesting that the behavior of $\Betti_2$ (red solid 
curve) is an almost perfect reflection of the behavior of $\Betti_0$ through the zero density level.
It documents the growing prominence of voids for negative and shrinking density threshold.
Here we find a maximum value for $\nu$ approximately equal to $- \sqrt{3}$. 

We note that we would get perfect symmetry if we replaced the number of components
by the number of gaps between them, which is $\Betti_0 - 1$.
As mentioned, $\Betti_2$ goes to zero rapidly as $\nu$ increases beyond $0$,
and $\Betti_0 - 1$ goes to zero rapidly as $\nu$ decreases below $0$.
However, at $\nu = 0$, both are small but clearly positive.
The expected number of components at $\nu = 0$ is thus small but larger than $1$,
with an expected number of voids that is precisely one less than for the components.

%%%%%%%%%%%%%%%%%%%%%%%%%%%%%%%%%%%%%%%%%%%%%%%%%%%%%%%%%%%%%%%%%%%%%%%%%%%%%%%%%
\subsection{Gaussian Fields Versus the Cosmic Web}
%%%%%%%%%%%%%%%%%%%%%%%%%%%%%%%%%%%%%%%%%%%%%%%%%%%%%%%%%%%%%%%%%%%%%%%%%%%%%%%%%

It is interesting to compare these results with the homology we find in the web-like configuration
of the evolved Universe.
Besides the disappearance of symmetry between high-density and low-density regions, reflected in 
a substantially different behavior of $\Betti_0$ and $\Betti_2$, there 
are a few additional differences as well as similarities. 

In evolving LCDM density fields, we find that $\Betti_1$ is almost always
in excess of $\Betti_2$: the number of tunnels in the Cosmic Web tends to be
several factors higher than the number of enclosed voids; see Figure~\ref{fig:bettilcdm}.
However, the advanced nonlinear mass distribution is marked by a substantially 
higher number of individual components than found in the Gaussian field. This partially 
reflects the difference of an alpha shape based analysis and one based on level sets 
defined for a fixed Gaussian filter radius. In the alpha shape analysis, the substructure 
of an intrinsically multiscale mass distribution emerging through an hierarchical 
is not lost. At small values of $\alpha$ one finds the small clumps that in a 
Gaussian filtered field have been removed. 

%%%%%%%%%%%%%%%%%%%%%%%%%%%%%%%%%%%%%%%%%%%%%%%%%%%%%%%%%%%%%%%%%%%%%%%%%%%%%%%%%%%%%%
\subsection{Genus and Homology}
%%%%%%%%%%%%%%%%%%%%%%%%%%%%%%%%%%%%%%%%%%%%%%%%%%%%%%%%%%%%%%%%%%%%%%%%%%%%%%%%%%%%%%

It is interesting to relate the results on the Betti numbers depending on a threshold value
with the analytically known expression for the expected \emph{reduced genus}\footnote{In 
section~\ref{sec:genus} we remarked that the genus $g$ in cosmological studies is slightly 
differently defined than the usual definition of the genus $G$ (eqn.~\ref{eq:genus}). The 
genus, $\genusalt$, in these studies has been defined as the number of holes minus the 
number of connected regions: $\genusalt = \genus - c$. In this review we distinguish between 
these definitions by referring to $g$ as the \emph{reduced genus}.} 
in a Gaussian random field \cite{gott1986,hamilton1986}.
This expression is
\begin{eqnarray}
  g (\nu)  &=&  - \frac{1}{8\pi^2} \left( \frac{{\langle k^2 \rangle}}{3} \right)^{3/2}
               (1 - \nu^2) e^{-\nu^2/2},
  \label{eq:gaussgenus}
\end{eqnarray}
where $\langle k^2 \rangle = \langle |\nabla f|^2 \rangle / \langle f^2 \rangle$.
%% where $\langle k^2 \rangle = \sigma^2_1 / \sigma^2_0$,
%% $\sigma_0 = \langle \rho^2 \rangle^{1/2}$, and
%% $\sigma_1 = \langle |\nabla\rho|^2 \rangle^{1/2}$.
Other than its amplitude, the shape does not depend on the power spectrum but
only on whether or not the field is Gaussian.
Recall from \eqref{eqn:genus-Betti} that the genus of an isodensity surface is directly related
to the Betti numbers of the enclosed superlevel sets, via the alternating sum
\begin{eqnarray}
  g (\nu)  &=&  - \Betti_0 (\nu) + \Betti_1 (\nu) - \Betti_2 (\nu) .
  \label{eq:genusbettinu}
\end{eqnarray}
We may therefore compare the outcome of our superlevel set study to the expected distribution for 
the genus in \eqref{eq:gaussgenus}.
In Figure~\ref{fig:gaussbetti}, the magenta dotted dashed line is the genus $g (\nu)$
computed from the alternating sum \eqref{eq:genusbettinu}.
It closely matches the predicted relation \eqref{eq:gaussgenus}; see \cite{park11}. 
It is most reassuring that the peaks of $\Betti_0$ and $\Betti_2$ are reached at 
the threshold value where, according to \eqref{eq:gaussgenus},
$g(\nu)$ has its two minima, at $\nu = - \sqrt{3}$ and $\nu = \sqrt{3}$. 

Having established the cosmologically crucial Gaussian basis of the Betti number analysis, 
a few more interesting findings follow from our analysis in \cite{park11}.
While the shape of the genus function is independent of the power spectrum,
it turns out that the Betti numbers do reflect the power spectrum.
In particular, for small absolute values of $\nu$, there are substantial 
differences between the Betti numbers in Gaussian fields with different power spectra.
This makes them potentially strong discriminators between cosmological scenarios.
On the other hand, the differences are perfectly symmetric between $\Betti_0 - 1$ and $\Betti_2$,
and $\Betti_1$ is symmetric itself.
The Betti numbers may therefore be useful for tracing non-Gaussianities in the primordial Universe,
a major point of interest in current cosmological studies. 

%%%%%%%%%%%%%%%%%%%%%%%%%%%%%%%%%%%%%%%%%%%%%%%%%%%%%%%%%%%%%%%%%%%%%%%%%%%%%%%%%%%%%%
\subsection{Gaussian Betti Correlations}
%%%%%%%%%%%%%%%%%%%%%%%%%%%%%%%%%%%%%%%%%%%%%%%%%%%%%%%%%%%%%%%%%%%%%%%%%%%%%%%%%%%%%%

Finally, \cite{park11} also assess the correlations in Gaussian fields between the various 
Betti numbers. They find that the Betti numbers are not independent of one another. 
For example, $\Betti_0$ near its maximum ($\nu=1.7$) is positively correlated with $\Betti_1$ at 
low threshold levels, while there are other levels where $\Betti_0$ and $\Betti_1$ are 
anti-correlated. 

The important implication is that we need to take into account the mutual dependence 
of the Betti numbers. This will be largely dependent on the nature of the density 
field. On the other hand, in general the correlation is not perfect. In other words, 
the Betti numbers contains complementary information on the topological structure 
of e.g. the cosmic mass distribution. 

%%%%%%%%%%%%%%%%%%%%%%%%%%%%%%%%%%%%%%%%%%%%%%%%%%%%%%%%%%%%%%%%%%%%%%%%%%%%%%%%%%
%%%%%%%%%%%%%%%%%%%%%%%%%%%%%%%%%%%%%%%%%%%%%%%%%%%%%%%%%%%%%%%%%%%%%%%%%%%%%%%%%%
\medskip
\section{Persistence}
\label{sec:persist}
%%%%%%%%%%%%%%%%%%%%%%%%%%%%%%%%%%%%%%%%%%%%%%%%%%%%%%%%%%%%%%%%%%%%%%%%%%%%%%%%%%
%%%%%%%%%%%%%%%%%%%%%%%%%%%%%%%%%%%%%%%%%%%%%%%%%%%%%%%%%%%%%%%%%%%%%%%%%%%%%%%%%%

The one outstanding issue we have not yet addressed systematically is the hierarchical 
substructure of the Cosmic Web data.  In standard practice, the multiscale nature
of the mass distribution tends to be investigated by means of user-imposed filtering.
Persistence rationalizes this approach by considering the range of filters at once,
and by making meaningful comparisons.  At the same time, it deepens the approach by
combining it with topological measurements: the homology groups and their ranks.

\begin{figure*}[h]
 \vskip -0.5truecm
 \begin{center}
  \includegraphics[bb=0 0 953 479,width=0.98\textwidth]{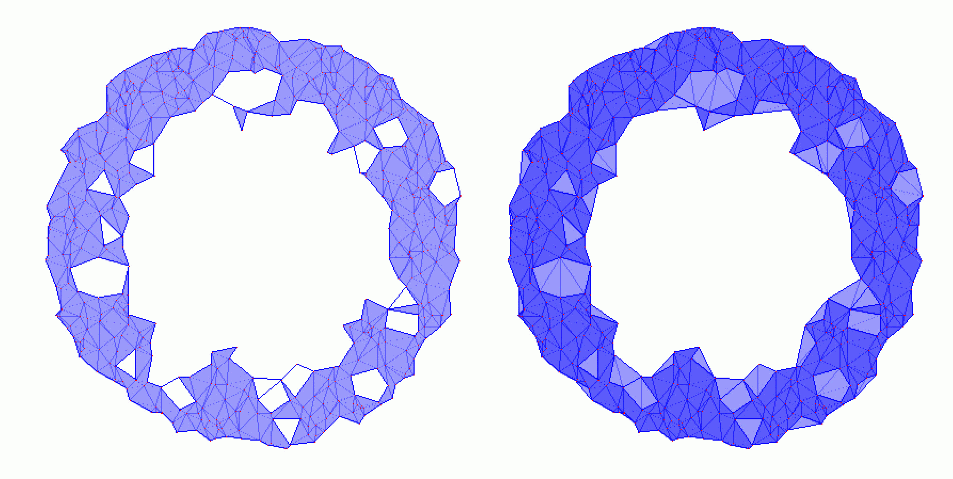} 
  \vskip -0.0truecm
  \caption{Illustrating the idea of persistence.  The alpha shapes of a set of points
    randomly chosen in a $2$-dimensional annulus.  Left: for a small value of $\alpha$,
    the alpha shape has $18$ holes.  Right: for a somewhat larger value of $\alpha$,
    the alpha shape has only two holes.  One of these holes is new, while the other
    already existed on the left, albeit in somewhat different form.
    We say the latter hole \emph{persists} over the entire interval delimited by
    the two alpha shapes.}
\label{fig:persistence}
\end{center}
\vskip -0.75truecm
\end{figure*}

Persistence entails the conceptual framework and language for separating scales of a 
spatial structure or density distribution.  It was introduced by Edelsbrunner,
Letscher and Zomorodian \cite{edelsbrunner2002}; see also the recent text on
computational topology \cite{edelsbrunner2010} of which persistent homology forms the core.
By separating the scales in a mass distribution, one may analyze and map the
topological hierarchy of the cosmos.  Within this context, substructures can be
separated by determining the range of scales over which they exist.
Here, we find a close link to Morse theory, in which the crucial new idea contributed
by persistence is the pairing of all critical points depending on a global
topological criterion.
A recent discussions of persistence in the cosmological context can be found
in the work of Sousbie et al. \cite{sousbie11a,sousbie11b}.

%%%%%%%%%%%%%%%%%%%%%%%%%%%%%%%%%%%%%%%%%%%%%%%%%%%%%%%%%%%%%%%%%%%%%%%%%%%%%%%%%%%%%%
\subsection{Persistence of Alpha Shapes}
%%%%%%%%%%%%%%%%%%%%%%%%%%%%%%%%%%%%%%%%%%%%%%%%%%%%%%%%%%%%%%%%%%%%%%%%%%%%%%%%%%%%%%

Alpha shapes provide the perfect context to illustrate the essential idea of persistence,
which is to follow components, tunnels, and voids over the entire range of the
parameter, $\alpha$.  The $2$-dimensional illustration in Figure~\ref{fig:persistence}
shows how the small scale holes in the left frame disappear when we increase the
parameter in the right frame.  This is not to say that the value of $\alpha$ chosen
on the right is perfect for the data set; indeed, we see a new, larger hole appear
that did not yet exist on the left.  The key point is that each feature is
\emph{born} at some value of $\alpha$, and \emph{dies} at another value of $\alpha$.
The interval between birth and death sets the position of the feature within
the structural hierarchy.  In particular, the length of the interval, that is,
the absolute difference between the values of $\alpha$ at the birth and at the death,
is called the \emph{persistence} of the feature.

\begin{figure*}[h]
 \vskip -0.5truecm
 \begin{center}
  \mbox{\hskip -0.5truecm\includegraphics[bb=0 0 1455 1121,width=1.06\textwidth]{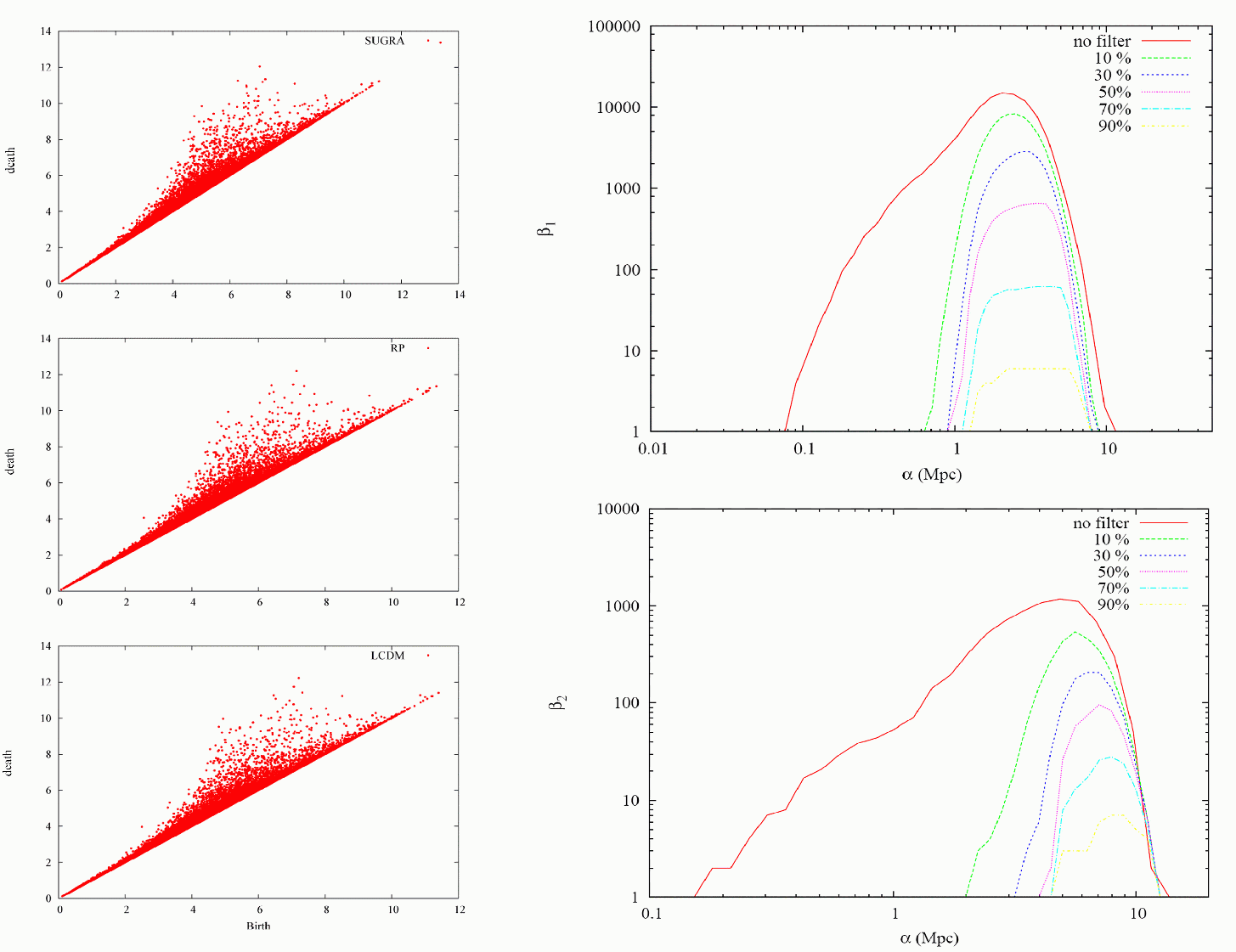}}
  \vskip -0.0truecm
  \caption{Persistent homology for the Cosmic Web simulations.
    Left:  the $1$-dimensional persistence diagrams of the LCDM simulation and of two
    quintessence simulations representing the RP and SUGRA dark energy models.
    Right:  the curves describing the evolution of the first and second Betti numbers
    for growing value of $\alpha$.  In addition to giving the Betti numbers, at every
    value of $\alpha$, we show how many of the counted features belong to the
    top $90\%$, $70\%$, $50\%$, $30\%$, $10\%$ most persistent features
    in the entire population.}
  \label{fig:persistlcdm}
 \end{center}
\vskip -0.5truecm
\end{figure*}

The full implementation of persistence in our topological study will be addressed
in future publications; e.g. \cite{pranav11}.  Here, we briefly address its effect
on the Betti number analysis.  We assess the matter distribution for three
cosmological scenarios: LCDM, and RP and SUGRA quintessence.
Each is used in a many-body simulation, generating a discrete point distribution
in $3$-dimensional space that reflects the intricate multi-scale organization of matter.
For each dataset, we compute the entire range of alpha shapes, encoded for efficiency
reasons in the Delaunay triangulation of the points, and we follow each cycle
through the sequence of alpha shapes, recording when it is born and when it dies.
Separating the results for different homological dimensions ($p = 0$ for components,
$p = 1$ for tunnels, and $p = 2$ for voids), we show the statistics for $p = 1$
in Figure~\ref{fig:persistlcdm}.
On the left, we see the $1$-dimensional persistence diagrams of the three datasets.
Each dot represents a $1$-cycle in the alpha shape filtration.  Its horizontal
coordinate gives the value of $\alpha$ at which the $1$-cycle is born, which happens
when its last edge is added to the alpha shape.  The vertical coordinate of the dot
gives the value of $\alpha$ at which the $1$-cycle dies, which happens when the
last triangle completes a membrane filling the tunnel formed by the cycle.

In all three diagrams, we see a substantial number of $1$-cycles that die shortly
after they are born.  They correspond to dots close to and right above the diagonal
line in the diagram.
This is typical for natural data, in which we see features appear and disappear
in rapid progression at all times.  This ``topological noise'' creates a deceivingly
complicated picture, obscuring the cleaner picture of more persistent features.
In our datasets, there are also a good number of $1$-cycles that remain for a while
after they are born.  They correspond to dots that are further up in the persistence
diagram, with the vertical distance to the diagonal representing the measured persistence.

%%%%%%%%%%%%%%%%%%%%%%%%%%%%%%%%%%%%%%%%%%%%%%%%%%%%%%%%%%%%%%%%%%%%%%%%%%%%%%%%%%%%%%
\subsection{Persistence and Scale}
%%%%%%%%%%%%%%%%%%%%%%%%%%%%%%%%%%%%%%%%%%%%%%%%%%%%%%%%%%%%%%%%%%%%%%%%%%%%%%%%%%%%%%

The promise of persistence is that it gives access to data on different scale levels.
This is particularly important for the Cosmic Web, where we observe structure
on almost every scale.  The goal would be to separate scales, or quantify them in
such a way that relations between scales become explicit.

To probe this aspect of persistence, we use a \emph{persistence parameter}, $\lambda$,
that restricts our attention to features of persistence at least $\lambda$.
In our experiment, we adjusted the parameter so that a specified percentage of the
features (dots) have persistence at least $\lambda$.  All dots with persistence less
than $\lambda$ are removed.  The effect on the Betti numbers of the LCDM simulation data
is illustrated in the two right frames of Figure~\ref{fig:persistlcdm}.
The top frame shows six evolutions of $\beta_1$ for increasing values of $\alpha$:
for $100\%$, $90\%$, $70\%$, $50\%$, $30\%$, and $10\%$ of the most persistent
$1$-cycles.  The bottom frame shows the same information for $\beta_2$.

The features with small persistence can be considered noise.  They may be reflections
of numerical inaccuracies in computations, artefacts of the data representation,
measuring errors, or genuine features that are too small for us to distinguish them
from noise.  In practical astrophysical circumstances, datasets are indeed beset
by a variety of artefacts.  As a result, the alpha shapes of discrete point
distributions, such as the galaxies in our local cosmic environment, will reflect the noise.
The induced irregularities in the number of components, tunnels, and voids do not
represent any real topological structure but nonetheless influence the Betti numbers
we collect.  Persistence diagrams offer the real opportunity to remove the noise
without introducing side-effects, and they may help in defining the usually
ill-specified noise distribution.

%%%%%%%%%%%%%%%%%%%%%%%%%%%%%%%%%%%%%%%%%%%%%%%%%%%%%%%%%%%%%%%%%%%%%%%%%%%%%%%%%%
%%%%%%%%%%%%%%%%%%%%%%%%%%%%%%%%%%%%%%%%%%%%%%%%%%%%%%%%%%%%%%%%%%%%%%%%%%%%%%%%%%
\medskip
\section{Conclusions and Prospects}
\label{sec:conclusions}
%%%%%%%%%%%%%%%%%%%%%%%%%%%%%%%%%%%%%%%%%%%%%%%%%%%%%%%%%%%%%%%%%%%%%%%%%%%%%%%%%%
%%%%%%%%%%%%%%%%%%%%%%%%%%%%%%%%%%%%%%%%%%%%%%%%%%%%%%%%%%%%%%%%%%%%%%%%%%%%%%%%%%

In this overview we have discussed and presented an analysis of the homology of 
cosmic mass distributions. We have argued that Betti numbers of density level sets 
and of the corresponding isodensity surfaces provide us with a far more complete 
description of the topology of web-like patterns in the megaparsec Universe than 
the well-known genus analysis or the topological and geometric instruments of 
Minkowski functionals. The genus of isodensity surfaces and the Euler characteristic, 
the main topological Minkowski functional, are directly related to Betti numbers 
and may be considered as lower-dimensional projections of the homological 
information content in terms of Betti numbers.

We have established the promise of alpha shapes for measuring the homology of the 
megaparsec galaxy distribution in the Cosmic Web. Alpha shapes are well-defined 
subsets of Delaunay tessellations, selected following a strictly defined 
scale parameter $\alpha$. These simplicial complexes constitute a filtration of the 
Delaunay tessellations, and are homotopy equivalent to the distance function field 
defined by the point distribution. Alpha shape analysis has the great 
advantage of being self-consistent and natural, involving shapes and surfaces that 
are entirely determined by the point distribution, independent of any artificial filtering. 

By studying the Betti numbers and several Minkowski functionals of a set of heuristic 
Voronoi clustering models, as well as dark energy cosmological scenarios (LCDM, 
and RP and SUGRA quintessence), we have illustrated the potential for exploiting 
the cosmological information contained in the Cosmic Web. We have analyzed the 
evolution of Betti numbers in an LCDM simulation, and compared the discriminative 
power of Betti curves for distinguishing between different dark energy models. 
In addition, we have addressed the significance of Betti numbers in the case of 
Gaussian random fields, as these primordial density fields represent the reference 
point for any further analysis of the subsequent cosmic structure evolution. 

\medskip
\noindent {\it Related Work}\\
The mathematical fundamentals of our project, including a full persistence analysis, will be extensively 
outlined and defined in the upcoming study of Pranav, Edelsbrunner et al. \cite{pranav11}. In addition to an 
evaluation on the basis of the simplicial alpha complexes, it also discusses and compares the homology analysis 
on the basis of isodensity surfaces, superlevel and sublevel sets, and their mutual relationship. This includes 
an evaluation of the results obtained on the basis of a piecewise linear density field reconstructions on Delaunay 
tessellations, following the DTFE formalism \cite{schaapwey2000,weyschaap2009,cautun11}. 

In related work, \cite{sousbie11a} and \cite{sousbie11b} recently published an impressive study on the topological 
analysis of the Cosmic Web within the context of the skeleton formalism \cite{sousbie2008,sousbie2008b,sousbie2009}. 
Their study used the DTFE density field reconstructions and incorporated the complexities of a 
full persistence analysis to trace filamentary features in the Cosmic Web.  

In two accompanying letters we will address cosmological applications of homology analysis. 
In \cite{weygaert11} the ability of Betti numbers to discriminate between different dark energy cosmological 
scenarios will be demonstrated. Crucial for furthering our understanding and appreciation of the significance of 
measured Betti numbers in a cosmological context, is to know their values in Gaussian random density fields. 
With cosmic structure emerging from a primordial Gaussian field, the homology of these fields forms a crucial 
reference point. In the study by Park et al. \cite{park11} we have analyzed this aspect in more detail.

%%%%%%%%%%%%%%%%%%%%%%%%%%%%%%%%%%%%%%%%%%%%%%%%%%%%%%%%%%%%%%%%%%%%%%%%%%%%%%%%%%%%%%%
\medskip
\section*{Acknowledgement}
%%%%%%%%%%%%%%%%%%%%%%%%%%%%%%%%%%%%%%%%%%%%%%%%%%%%%%%%%%%%%%%%%%%%%%%%%%%%%%%%%%%%%%%

We thank Dmitriy Morozov for providing us with his code for computing persistence diagrams 
and Michael Kerber for his homology code for computing Betti numbers of grid represented 
manifolds. In addition, we thank Bartosz Borucki for his permission to use figure 1. We also 
acknowledge many interesting and enjoyable discussions with Mathijs Wintraecken. 
In particular, we wish to express our gratitude to the editor, Mir Abolfazl Mostafevi, 
for his help and patience at the finishing stage of this project. Part of this project 
was carried out within the context of the CG Learning project. The project CG Learning 
acknowledges the financial support of the Future and Emerging Technologies (FET) programme within 
the Seventh Framework Programme for Research of the European Commission, under FET-Open grant 
number: 255827.

% trigger a \newpage just before the given reference
% number - used to balance the columns on the last page
% adjust value as needed - may need to be readjusted if
% the document is modified later
%\IEEEtriggeratref{8}
% The "triggered" command can be changed if desired:
%\IEEEtriggercmd{\enlargethispage{-5in}}

% references section

% can use a bibliography generated by BibTeX as a .bbl file
% BibTeX documentation can be easily obtained at:
% http://www.ctan.org/tex-archive/biblio/bibtex/contrib/doc/
% The IEEEtran BibTeX style support page is at:
% http://www.michaelshell.org/tex/ieeetran/bibtex/
%\bibliographystyle{IEEEtran}
% argument is your BibTeX string definitions and bibliography database(s)
%\bibliography{IEEEabrv,../bib/paper}
%
% <OR> manually copy in the resultant .bbl file
% set second argument of \begin to the number of references
% (used to reserve space for the reference number labels box)
% can use a bibliography generated by BibTeX as a .bbl file
% BibTeX documentation can be easily obtained at:
% http://www.ctan.org/tex-archive/biblio/bibtex/contrib/doc/
% The IEEEtran BibTeX style support page is at:
% http://www.michaelshell.org/tex/ieeetran/bibtex/
\bibliographystyle{splncs03}
%\bibliography{cosmo_alpha_betti}
\bibliography{Rien_Weygaert_cosmoalphabetti_arxiv}

% that's all folks
\end{document}